\documentclass[aps,superscriptaddress,amsmath,amsfonts,amssymb,12pt,%
tightenlines,%
nofootinbib,
eqsecnum]{revtex4}

\usepackage{graphicx}
\usepackage{bm}
\usepackage{empheq}
\usepackage{simplewick}

\DeclareMathOperator{\tr}{tr}
\DeclareMathOperator{\sinc}{sinc}

\DeclareMathOperator{\diag}{diag}

\DeclareMathOperator{\Realpart}{Re}

\def\tzero{0}

\def\1{\mat 1}

\def\inttzT{\int_{\tzero}^T}
\def\infint{\int_{-\infty}^{\infty}}
\def\halfint{\int_0^\infty}
\def\intsym{\int_{-T/2}^{T/2}}

\def\vec#1{\bm{\mathrm{#1}}}

\def\lcolon{\mathopen{:}}
\def\rcolon{\mathclose{:}}

\def\avg#1{\left\langle {#1} \right\rangle}
\def\abs#1{\left\lvert{#1}\right\rvert}

\def\hat{}
\def\laser{\mathrm{L}}
\def\atom{\mathrm{A}}
\def\boltz{\mathrm{B}}

\def\timeord{{\mathcal T}}

\def\mat#1{\mathsf{#1}}

\def\hadam{\circ}
\def\negspace{\!}
\def\lsub#1#2{{\vphantom{#1}}_{#2} \negspace {#1}}
\def\rsub#1#2{{#1} \negspace {\vphantom{#1}}_{#2}}

\def\bra#1{\left\langle {#1} \right\rvert}
\def\ket#1{\left\lvert {#1} \right\rangle}

\def\ketsub#1#2{\rsub {\ket{#1}} {#2}}
\def\brasub#1#2{\lsub {\bra{#1}} {#2}}

\def\mbra#1{\brasub{#1} m}
\def\nbra#1{\brasub{#1} n}
\def\mket#1{\ketsub{#1} m}
\def\nket#1{\ketsub{#1} n}

\def\outprod#1#2{\ket {#1}\!\bra {#2}}

\def\outprodsubsub#1#2#3#4{\ketsub {#1}{#3} \brasub{#2}{#4}}

\def\noutprod#1#2{\outprodsubsub{#1}{#2}n n}
\def\moutprod#1#2{\outprodsubsub{#1}{#2}m m}

\begin{document}

\title{Spatial correlation functions for the collective degrees of freedom of many trapped ions}
 
\author{Nicolas~C.~Menicucci}
\email{nmen@princeton.edu}
\affiliation{Department of Physics, Princeton University, Princeton, NJ 08544, USA}
\affiliation{Department of Physics, The University of Queensland, Brisbane, Queensland 4072, Australia}

\author{G.~J.~Milburn}
\affiliation{Department of Physics, The University of Queensland, Brisbane, Queensland 4072, Australia}

\date{\today}

\begin{abstract}
Spatial correlation functions provide a glimpse into the quantum correlations within a quantum system.  Ions in a linear trap collectively form a nonuniform, discretized background on which a scalar field of phonons propagates.  Trapped ions have the experimental advantage of each having their own ``built-in'' motional detector: electronic states that can be coupled, via an external laser, to the ion's vibrational motion.  The post-interaction electronic state can be read out with high efficiency, giving a stochastic measurement record whose classical correlations reflect the quantum correlations of the ions' collective vibrational state.  Here we calculate this general result, then we discuss the long detection-time limit and specialize to Gaussian states, and finally we compare the results for thermal versus squeezed states.
\end{abstract}

\maketitle

%%%%%%%%%%%%%%%%%%%%%%%%%%%%%%%%%%
% Introduction
%%%%%%%%%%%%%%%%%%%%%%%%%%%%%%%%%%

\section{Introduction}
\label{sec:intro}

In contexts as diverse as cosmology, quantum optics and condensed matter physics, spatial correlation functions provide experimental access to important features of spatially distributed quantum systems. In a cosmological setting, spatial correlations in the temperature distribution of the cosmic microwave background provide direct access to the fluctuations of a primordial quantum field during inflation. Details of these spatial correlations provide direct and detailed tests of inflationary cosmological models~\cite{Bernui2005}. For example, Grishchuk~\cite{Grishchuk1996} has suggested a model in which anisotropies reflect the underlying statistics of  squeezed vacuum states.  In condensed matter physics, the change in the length scale of correlations of the XY model in two dimensions can reveal a Kosterlitz-Thouless transition and vortex phases~\cite{Posazhennikova2006,Wen2004}.  The spatial correlation functions of a Bose-Einstein condensate, as reveled by light scattering from a freely expanding condensate reveal details of the quantum state of the condensate before expansion~\cite{Hellweg2003}. As one example of this, mean field energy of a condensate is a direct measure of the second order correlation function~\cite{Burt1997}. The atom loss-rate due to three-body recombination in a BEC is directly related to the probability of finding three atoms close to each other and can therefore act as a probe of a third-order correlation function~\cite{Kagan1985}. In this paper we show that similar two-point spatial correlation functions can be used to probe the collective vibrational state for a string of trapped ions~\cite{Leibfried2003}. Our work is related to that of  Franke-Arnold~\cite{Franke-Arnold2003}, which considers the spatial coherence properties of just two harmonically trapped particles.

Using external lasers, it is possible to couple an internal electronic transition to the vibrational motion of the ion~\cite{Monroe1995}. Indeed this is how laser cooling of multiple ions to the vibrational ground state of one or more of their collective normal modes of vibration~\cite{James1998}.  Simple quantum information processing tasks can be realised by coupling internal states through collective motional degrees of freedom~\cite{Leibfried2003a,Schmidt-Kaler2003}.  In this paper, we show how the ability to couple internal and vibrational degrees of freedom, plus the ability to efficiently read out the internal electronic state using a fluorescence shelving scheme~\cite{Leibfried2003}, provides a path to experimentally measuring spatial correlation functions for the collective vibrational degrees of freedom.  This provides a discrete analogue of spatial correlation functions for a scalar quantum field.

The scheme has two components: (1)~external lasers weakly couple the motion of two distinct ions to their local displacement from equilibrium; subsequently, (2)~external readout lasers are used to probe the electronic state of all the ions. In the second stage, some ions will ``light up'' (made more precise below), while others remain dark.  Thus, for $N$~ions, we get an $N$-element stochastic binary string $(e_1,e_2,\dotsc, e_N)$, with $e_m=1$~if an ion lights up and $e_m=0$~otherwise.  This is the measurement record, with the ordering of the string corresponding exactly with the position ordering of the ions in the trap.  We can then define the empirical two-point correlation function~$P_{mn}$ as the classical average $P_{mn}=\mathcal E[e_m e_n]$.  Our objective in this paper is to relate this classical correlation function to the quantum mechanical correlations of the ion displacement amplitudes (e.g.,\ terms such as~$\avg{\hat{q}_m \hat{q}_{n}}$) and, furthermore, to determine characteristic experimental signatures of different collective quantum states of motion.

%%%%%%%%%%%%%%%%%%%%%%%%%%%%%%%%%%
% Spatial Correlation Functions
%%%%%%%%%%%%%%%%%%%%%%%%%%%%%%%%%%

\section{Spatial Correlation Functions}
\label{sec:spatial}

We summarise the standard results for the spatial correlation functions of a scalar field (see, for example, Ref.~\cite{Naraschewski1999}).  A scalar field~$\hat \phi (\vec x,t)$ may be expanded as
\begin{equation}
	\hat{\phi}(\vec x,t) = \sum_{\vec k} a_{\vec k} u_{\vec k}(\vec x,t)+a_{\vec k}^\dagger u_{\vec k}^*(\vec x,t)\;,
\end{equation}
where $u_{\vec k}(\vec x,t)$~is a positive-frequency mode function, and $u_{\vec k}^*(\vec x,t)$~is its negative-frequency counterpart.  In the commonly used case of plane waves (with box normalization), we can define the positive- and negative-frequency components of~$\phi(\vec x,t)$ as
\begin{subequations}
\label{eq:posnegfreq}
\begin{align}
	\psi(\vec x,t) &\coloneqq \frac{1}{\sqrt{V}}\sum_{\vec k} a_{\vec k} e^{+i(\vec k \cdot \vec x - \omega_k t)} \\
	\text{and} \qquad \psi^\dag(\vec x,t) &\coloneqq \frac{1}{\sqrt{V}} \sum_{\vec k} a^\dag_{\vec k} e^{-i(\vec k \cdot \vec x - \omega_k t)}\;,
\end{align}
\end{subequations}
respectively.  In the continuum limit, these operators satisfy the standard boson equal-time commutation relations:
\begin{align}
\label{eq:psicommute}
	[ \psi(\vec x,t), \psi^\dag(\vec x',t) ] = \delta(\vec x - \vec x')\;.
\end{align}
With all of the time-dependence in the exponential~$e^{\pm i \omega_k t}$, we can change to the Schr\"odinger picture simply by simply removing this piece (which is equivalent to evaluation at~$t=0$).

Now in the Schr\"odinger picture, we define the first- and second-order normally ordered spatial correlation functions,
\begin{subequations}
\begin{align}
	G^{(1)}(\vec{x},\vec{x}^\prime) & \coloneqq \avg{\hat{\psi}^\dagger(\vec{x})\hat{\psi}(\vec{x}^\prime)} \\
	\text{and} \qquad G^{(2)}(\vec{x},\vec{x}^\prime) & \coloneqq \avg{\hat{\psi}^\dagger(\vec{x})\hat{\psi}^\dagger(\vec{x}^\prime)\hat{\psi}(\vec{x}^\prime)\hat{\psi}(\vec{x})}\;.
\end{align}
\end{subequations}
We also define the normalized correlation functions,
\begin{subequations}
\label{eq:normcorr}
\begin{align}
	g^{(1)}(\vec{x},\vec{x}^\prime) & \coloneqq \frac{G^{(1)}(\vec{x},\vec{x}^\prime)}{\sqrt{G^{(1)}(\vec{x},\vec{x})}\sqrt{G^{(1)}(\vec{x}^\prime,\vec{x}^\prime)}}\\
	\text{and} \qquad g^{(2)}(\vec{x},\vec{x}^\prime) & \coloneqq \frac{G^{(2)}(\vec{x},\vec{x}^\prime)}{G^{(1)}(\vec{x},\vec{x})G^{(1)}(\vec{x}^\prime,\vec{x}^\prime)}\;.
\end{align}
\end{subequations}
If each mode is independently excited to a thermal state, with no mode-mode correlations, we can show that~\cite{Naraschewski1999}
\begin{equation}
	g^{(2)}(\vec{x},\vec{x}^\prime)=1+ \abs{g^{(1)}(\vec{x},\vec{x}^\prime)}^2\;,
\end{equation}
where the second term represents bosonic bunching. In the continuum limit,
\begin{equation}
	g^{(1)}(\vec{x},\vec{x}^\prime)=\frac{1}{N}\int d^3\vec{k}\ n(\vec{k})e^{i\vec{k}\cdot(\vec{x}-\vec{x}^\prime)}\;,
\end{equation}
where $n(\vec{k})$ represents the thermal occupation number of mode~$\vec{k}$ and $N$~is the total occupation number, given by
\begin{equation}
	N \coloneqq \int d^3\vec{k}\  n(\vec{k})\;.
\end{equation}
It is apparent that $g^{(2)}(\vec{x},\vec{x}^\prime)$, as a function of $(\vec{x}-\vec{x}')$, starts at a value of~2 and decreases to~1. How fast it decreases is a measure of the range of spatial correlation functions and depends on the $\vec k$-dependance of $n(\vec{k})$.

The example of the thermal states is a special case of a general result for all classical states of the field for which
\begin{equation}
g^{(2)}(\vec{x},\vec{x})\geq g^{(2)}(\vec{x},\vec{x}+\vec{y})
\end{equation}
Nonclassical states violate this inequality. The particular example of squeezed light has been studied in some detail~\cite{Kolobov1999}.  The effect of the squeezing is to induce spatial correlations that modulate an effective thermal background density.  In some cases this reduces the correlation function $G^{(2)}(\vec x,\vec x')$ \emph{below} the thermal value of~2, a phenomenon related to photon antibunching in the optical case~\cite{Nogueira2001}.  It can also increase it above~2.  The ability to change the first and second order correlation functions in this way is what lies behind the new field of quantum imaging~\cite{Kolobov2000}.  In Section~\ref{sec:examples}, we will contrast thermal and squeezed states in the case of vibrations in a linear ion trap and show that the latter have a similar nonclassical signature.

%%%%%%%%%%%%%%%%%%%%%%%%%%%%%%%%%%
% Measurement of Ion Trap Spatial Correlations
%%%%%%%%%%%%%%%%%%%%%%%%%%%%%%%%%%

\section{Measurement of Ion Trap Spatial Correlations}
\label{sec:measurement}

%
% Normal modes of vibration
%

\subsection{Normal modes of vibration}

Linearizing the potential created by the overall harmonic potential due to the trap electrodes and the mutual Coulomb repulsion between the ions, we can model the collective motion of the ions in a linear trap as a collection of coupled harmonic oscillators.  With a coupling matrix~$\mat A$ obtained from the linearized combination of these potentials, the Hamiltonian is given by~\cite{James1998}
\begin{align}
\label{eq:H0local}
	H_0 = \frac {1} {2 M} {\vec p^T \vec p} + \frac {M \nu^2} {2} \vec q^T \mat A \vec q \;,
\end{align}
where $\vec q = (q_1, \dotsc, q_N)^T$~is a column vector of operators corresponding to the displacement of each ion from its equilibrium position, $\vec p = (p_1, \dotsc, p_N)^T$~is a column vector of corresponding momenta, $M$~is the mass of each ion, and $\nu$~is the effective harmonic trap frequency provided by the trap electrodes (typically, $\nu \sim \text{a few~MHz}$~\cite{Leibfried2003}).  The coupling matrix~$\mat A$ is symmetric and positive-definite.  Thus, we can diagonalize it:
\begin{align}
\label{eq:Adiag}
	\mat A = \mat B^T \mat \Lambda \mat B\;,
\end{align}
where $\mat \Lambda = \diag(\mu_1, \dotsc, \mu_N)$~is a diagonal matrix of positive eigenvalues, arranged in ascending order.  The orthogonal matrix~$\mat B$ defines the transformation to normal-mode coordinates~$\vec Q = \mat B \vec q$ and momenta~$\vec P = \mat B \vec p$.  Equation~\eqref{eq:H0local} can be rewritten in these coordinates as
\begin{align}
	H_0 &= \frac {1} {2 M} {\vec P^T \vec P} + \frac {M \nu^2} {2} \vec Q^T \mat \Lambda \vec Q \nonumber \\
	&= \sum_{p=1}^N \frac {P_p^2} {2 M} + \frac {M \nu_p^2} {2} Q_p^2 \nonumber \\
	&= \sum_p \hbar \nu_p\, a_p^\dag a_p\;,
\end{align}
where
\begin{align}
\label{eq:nup}
	\nu_p \coloneqq \nu \sqrt{\mu_p}
\end{align}
is the oscillation frequency of normal mode~$p$, and (since the normal modes oscillate independently) we have diagonalized this Hamiltonian using the standard prescription for a set of independent harmonic oscillators, ignoring the zero-point energy.  The normal-mode raising and lowering operators satisfy
\begin{align}
\label{eq:normalmodecommute}
	[a_p, a_{p'}^\dag] = \delta_{pp'}\;,
\end{align}
as is appropriate for independent bosonic modes.

The local oscillations~$q_m$ about the ions' equilibrium positions are given, in the interaction picture, by~\cite{James1998}
\begin{align}
\label{eq:qoftdef}
	q_m(t) = \sum_{p=1}^N \sqrt{ \frac {\hbar} {2 M \nu_p} } b_m^{(p)} (a_p e^{-i \nu_p t} + a_p^\dag e^{i \nu_p t})\;,
\end{align}
where $m \in \{1, \dotsc, N\}$~labels the ion, and $b_m^{(p)}$ is an entry of~$\mat B$, which defines the spatial mode functions of the normal mode.\protect\footnote{Prefactor and phase conventions for~$q_m$ vary in the literature.}  We can define a unitless version of this local displacement oeprator, as well:
\begin{align}
\label{eq:phidef}
	\hat \phi_m (t) \coloneqq \sum_{p=1}^N \frac {b_m^{(p)}} {{\mu_p}^{1/4}} ( e^{-i\nu_p t} \hat a_p + e^{i\nu_p t} \hat a_p^\dag )\;.
\end{align}
The connection between the two is given by
\begin{align}
	q_m(t) = \sqrt{ \frac {\hbar} {2 M \nu} } \phi_m(t) \;.
\end{align}
The positive- and negative-frequency components of~$\phi_m(t)$ may be defined analogously to those of Eqs.~\eqref{eq:posnegfreq}:
\begin{align}
\label{eq:ionposnegfreq}
	\psi_m(t) &\coloneqq \sum_{p=1}^N \frac {b_m^{(p)}} {{\mu_p}^{1/4}} e^{-i\nu_p t} \hat a_p \nonumber \\
	\text{and} \qquad \psi^\dag_m(t) &\coloneqq \sum_{p=1}^N \frac {b_m^{(p)}} {{\mu_p}^{1/4}} e^{+i\nu_p t} \hat a_p^\dag \;.
\end{align}
Because the coupling matrix~$\mat A$ only approximates a simple ``balls on a string'' coupling (characteristic of a bosonic field), the equal-time commutation relations for these operators only approximate those of a boson field (compare with Eq.~\eqref{eq:psicommute}):
\begin{align}
\label{eq:psicommuteion}
	[ \psi_m(t), \psi^\dag_n(t) ] &= \sum_{p=1}^N \frac {b_m^{(p)} b_n^{(p)}} {\sqrt{\mu_p}} = \left( \mat A^{-1/2} \right)_{mn}\;.
\end{align}
Still, the fact that the ions' \emph{normal-mode} operators commute properly, as in Eq.~\eqref{eq:normalmodecommute}, means that we don't need to worry about this, as long as we do our calculations using the normal modes explicitly.

%
% Laser-induced coupling of vibrational and electronic states
%

\subsection{Laser-induced coupling of vibrational and electronic states}

The abstract relations between the field spatial correlation functions described in Section~\ref{sec:spatial} ultimately are manifest in the observed correlations in spatially distributed detectors of some kind.  In the case of quantum optics, for example, one imagines that the field falls on a photodetector array. Each element of the array produces a photocurrent~$I(x,t)$ indexed by the position of that photodetector in the array. We can then look at cross-correlations between photo currents from different detectors in the array. It is the the objective of measurement theory to determine how those spatially dependent current-current correlation functions are determined at a fundamental level by the field spatial correlation functions themselves. Explicit formulae are given in Kolobov~\cite{Kolobov1999} using the standard quantum optics theory of photodetection. 

Our measurement model is different from that considered in quantum optics and so we now explicitly make the connection between the observations and the underlying field correlation functions for the motion of the ions in the trap. The feature of our model is that the internal electronic state of each ion can be turned into a local detector for the displacement of that ion.  To achieve this for a given ion, its internal state must become correlated with its linear displacement from equilibrium.

This correlation is provided by an external laser, which is used to drive an electronic transition between two meta-stable electronic levels~$\ket g$ and~$\ket e$, separated in energy by~$\hbar\omega_\atom$~\cite{Wallentowitz1996,Leibfried2003,James1998}.  The interaction between an external classical laser field and the $m$th~ion is described, in the dipole and rotating-wave approximation, by the interaction-picture Hamiltonian~\cite{Leibfried2003,James1998}
\begin{align}
\label{eq:HIdef}
	H_I^{(m)} = -i\hbar\Omega_0 \left[\sigma_+^{(m)}(t)e^{ik\cos\theta q_m(t)}-\sigma_-^{(m)}(t)e^{-ik\cos\theta q_m(t)}\right]\;,
\end{align}
where $\Omega_0$~is the Rabi frequency for the laser-atom interaction (typically, $\Omega_0 \sim 100~\text{kHz}$~\cite{Leibfried2003}), $k$~is the magnitude of the laser's wave vector~$\vec k$, which makes an angle~$\theta$ with the trap axis, $q_m(t)$~is the interaction-picture position operator for the $m$th~ion, and $\sigma_\pm^{(m)}(t)$ are its interaction-picture electronic raising and lowering operators.  Explicitly,
\begin{equation}
\label{eq:sigmapmdef}
	\sigma_\pm^{(m)}(t) = e^{\pm i \Delta t} \sigma_\pm^{(m)}\;,
\end{equation}
where~$\sigma_+^{(m)}=\moutprod e g$ and~$\sigma_-^{(m)}=\moutprod g e$, and
\begin{equation}
\label{ion:eq:Deltadef}
	\Delta=\omega_\atom-\omega_\laser
\end{equation}
is the detuning of the laser below the atomic transition.  The size of the rms~fluctuation in~$q_m$ as compared to the wavelength of the laser is measured by the \emph{Lamb-Dicke parameter}
\begin{align}
\label{eq:etadef}
	\eta \coloneqq \sqrt{ \frac {\hbar k^2 \cos^2 \theta } {2 M \nu} } \sim \frac {\Delta x^{(m)}_{\text{rms}}} {\lambda_\laser} (2 \pi \cos \theta) \;,
\end{align}
where $\sqrt{\hbar/2 M\nu}$~is the rms~fluctuation of the center-of-mass mode of the ions in the ground state.  This quantity is representative of the overall rms~fluctuations of the $m$th ion, denoted~$\Delta x^{(m)}_{\text{rms}}$, since the frequencies~$\nu_p$ for all higher normal modes of the ions remain within an order of magnitude of~$\nu$ for small numbers of ions (up to $ N \sim 10$), realistic for current experiments~\cite{James1998}.  Typical values of the Lamb-Dicke parameter are~\mbox{$\eta \sim$ 0.01~to~0.1}~\cite{Walls2008}.  When~$\eta \ll 1$, the so-called ``Lamb-Dicke limit,'' the ion is well localized with respect to the wavelength of the laser, and we can expand the exponentials in Eq.~\eqref{eq:HIdef} to first order in~$\eta$, which is equivalent to first order in~$k \cos \theta\, q_m(t)$, giving
\begin{align}
\label{eq:HIexpand}
	H_I^{(m)}(t) &\simeq %-i\hbar\Omega_0 \Bigl[\sigma_+(t) \bigl[1 + ik\cos\theta\, q(t) \bigr] -\sigma_-(t) \bigl[1 - ik\cos\theta\, q(t) \bigr] \Bigr] \nonumber \\
%	&= \hbar\Omega_0 \Bigl[-i \sigma_+(t) +i \sigma_-(t) \Bigr] + \hbar\Omega_0 k\cos\theta\, q(t) \Bigl[ \sigma_+(t)  + \sigma_-(t) \Bigr] \nonumber \\
%	&= 
	\underbrace{\hbar\Omega_0 \sigma_y^{(m)}(t)}_{\text{carrier}} + \underbrace{\hbar\Omega_0 k\cos\theta\, q_m(t) \sigma_x^{(m)}(t)}_{\text{sideband}}\;,
\end{align}
where~$\sigma_x^{(m)}(t) = e^{+i \Delta t} \sigma_+^{(m)} + e^{- i \Delta t} \sigma_-^{(m)}$, and~$\sigma_y^{(m)}(t) = -i e^{+i \Delta t} \sigma_+^{(m)} +i e^{- i \Delta t} \sigma_-^{(m)}$.  The first term corresponds to excitation of the transition directly by the laser, while the second couples the atomic transition to vibrational motion.

The ``carrier'' term is resonant when~$\Delta = 0$ and corresponds to direct excitation of the atomic transition, which does not couple to the vibrational motion at all.  For sufficiently long-time detection, another rotating-wave approximation may be made in which only resonant (nonoscillatory) terms are kept.  In this case, if the carrier transition is sufficiently off-resonant ($\Delta \ne 0$), then it can be neglected, leaving
\begin{align}
\label{eq:HIsideband}
	H_I^{(m)}(t) &\xrightarrow[\text{($\Delta \ne 0$)}]{\text{sideband}} \hbar\Omega_0 \eta \sum_{r=1}^N \frac {b_m^{(r)}} {\mu_r^{1/4}} (e^{-i \nu_r t} a_r + e^{i \nu_r t} a_r^\dag) ( e^{ i \Delta t} \sigma_+^{(m)} + e^{- i \Delta t} \sigma_-^{(m)})\;.
\end{align}
This remaining ``sideband'' term can be used to couple the atomic transition to the vibrational motion through a judicious choice of detuning, $\Delta = \pm \nu_p$, for some normal mode frequency~$\nu_p$.  The first red sideband transition is obtained by setting $\Delta = \nu_p$; that is, the laser is detuned one unit of vibrational energy \emph{below} (to the red of) the atomic transition.  In this case, the resonant terms are
\begin{align}
\label{eq:HIredmodep}
	H_I^{(m)}(t) &\xrightarrow[\text{($\Delta = \nu_p$)}]{\text{red sideband, mode~$p$}} \hbar\Omega_0 \eta \frac {b_m^{(p)}} {\mu_p^{1/4}} (a_p \sigma_+^{(m)} + a_p^\dag \sigma_-^{(m)})\;.
\end{align}
This is of Jaynes-Cummings form~\cite{Jaynes1963,Leibfried2003}, corresponding to excitation of the atomic transition with energy gap~$\hbar \omega_\atom = \hbar \omega_\laser + \hbar \nu_p$ upon absorption of one laser photon at energy~$\hbar \omega_\laser$, along with absorption of one vibrational phonon at energy~$\hbar \nu_p$.  Detuning the laser above (to the blue of) the atomic transition by one unit of vibrational energy~($\Delta = -\nu_p$, for some mode~$p$) generates the blue sideband transition, which has resonant terms
\begin{align}
\label{eq:HIbluemodep}
	H_I^{(m)}(t) &\xrightarrow[\text{($\Delta = -\nu_p$)}]{\text{blue sideband, mode~$p$}} \hbar\Omega_0 \eta \frac {b_m^{(p)}} {\mu_p^{1/4}} (a_p^\dag \sigma_+^{(m)} + a_p \sigma_-^{(m)})\;.
\end{align}
This corresponding to atomic excitation of the transition at energy~$\hbar \omega_\atom = \hbar \omega_\laser - \hbar \nu_p$ upon absorption of one laser photon at energy~$\hbar \omega_\laser$, along with \emph{emission} of one vibrational phonon at energy~$\hbar \nu_p$.

The first red-sideband transition uses the ion itself as its own detector of motional quanta and this configuration comprises the first half of our two-stage detection model.  As long as the coupling strength~$\Omega_0 \eta$ is weak enough, no more than a single $\ket g \to \ket e$ transition will be excited for a given ion in the time during which the laser coupling is active.  Although the approximations above simplify the interaction Hamiltonians and provide a means to understand the physical mechanisms at work, for finite detection times the full form of the interaction may be needed.  As such, we will retain the ``sideband'' term of Eq.~\eqref{eq:HIexpand} as the interaction Hamiltonian, which has the form of a De~Witt monopole coupling~\cite{DeWitt1979}, in order to allow for greater applicability of our results.  In the examples given in Section~\ref{sec:examples}, we will apply the long detection-time approximation at the end of the calculations.

%
% Excitation probabilities and correlation functions
%

\subsection{Excitation probabilities and correlation functions}

In the interaction picture, the full interaction Hamiltonian is
\begin{align}
\label{eq:HsubI}
	H_I(t) \coloneqq \hbar \Omega_0 \eta {\sum_m}'  \hat \phi_m (t) \sigma_x^{(m)}(t)\;,
\end{align}
where the prime on the sum indicates that only the ions being addressed by interaction lasers are included, and the unitless displacement operator~$\phi_m(t)$ is defined in Eq.~\eqref{eq:phidef}.  The time evolution operator may be written formally as
\begin{align}
\label{eq:Udef}
	U(T) \coloneqq \timeord \left\{ \exp \left[ \frac{-i}{\hbar} \inttzT dt\, H_I(t) \right] \right\}\;,
\end{align}
where $\timeord$ is the Dyson time-ordering symbol.  The time-ordered exponential is defined by the time-ordering of its Taylor-series expansion, the first few terms of which are written here:
\begin{align}
\label{eq:Uexpand}
	U(T) = \1 + \frac{-i}{\hbar} \inttzT dt\, H_I(t) + \frac {1}{2!} \left( \frac{-i}{\hbar} \right)^2 \inttzT dt_1 \inttzT dt_2\, \timeord\{H_I(t_1) H_I(t_2)\} + \dotsb\;.
\end{align}
We will use these three terms in what follows.

As discussed in the introduction, we wish to calculate correlation functions for simultaneous detected excitations.  Detecting the excitation of a given ion~$m$ corresponds to measuring the projector
\begin{align}
\label{eq:calPm}
	\mathcal P_m &\coloneqq \moutprod e e \;,
\end{align}
where $\mket e$ is the excited electronic state of the $m$th ion.  $\mathcal P_m$ acts trivially on all other electronic states and on all vibrational modes. In the experiment, the electronic state of the ion is determined using the technique of fluorescence on a cycling transition~\cite{Leibfried2003}. The excited state $\mket e$ is caused to make a dipole allowed transition to another auxiliary level which then decays back to the state $\mket e$ through spontaneous emission. If this transition is saturated, a very large fluorescent photon flux is easily detected. The measurement very nearly approaches a projection measurement of the operator~$\mathcal P_m$ with an efficiency greater than 99\%. 

\begin{figure}
\includegraphics[width=.6 \columnwidth]{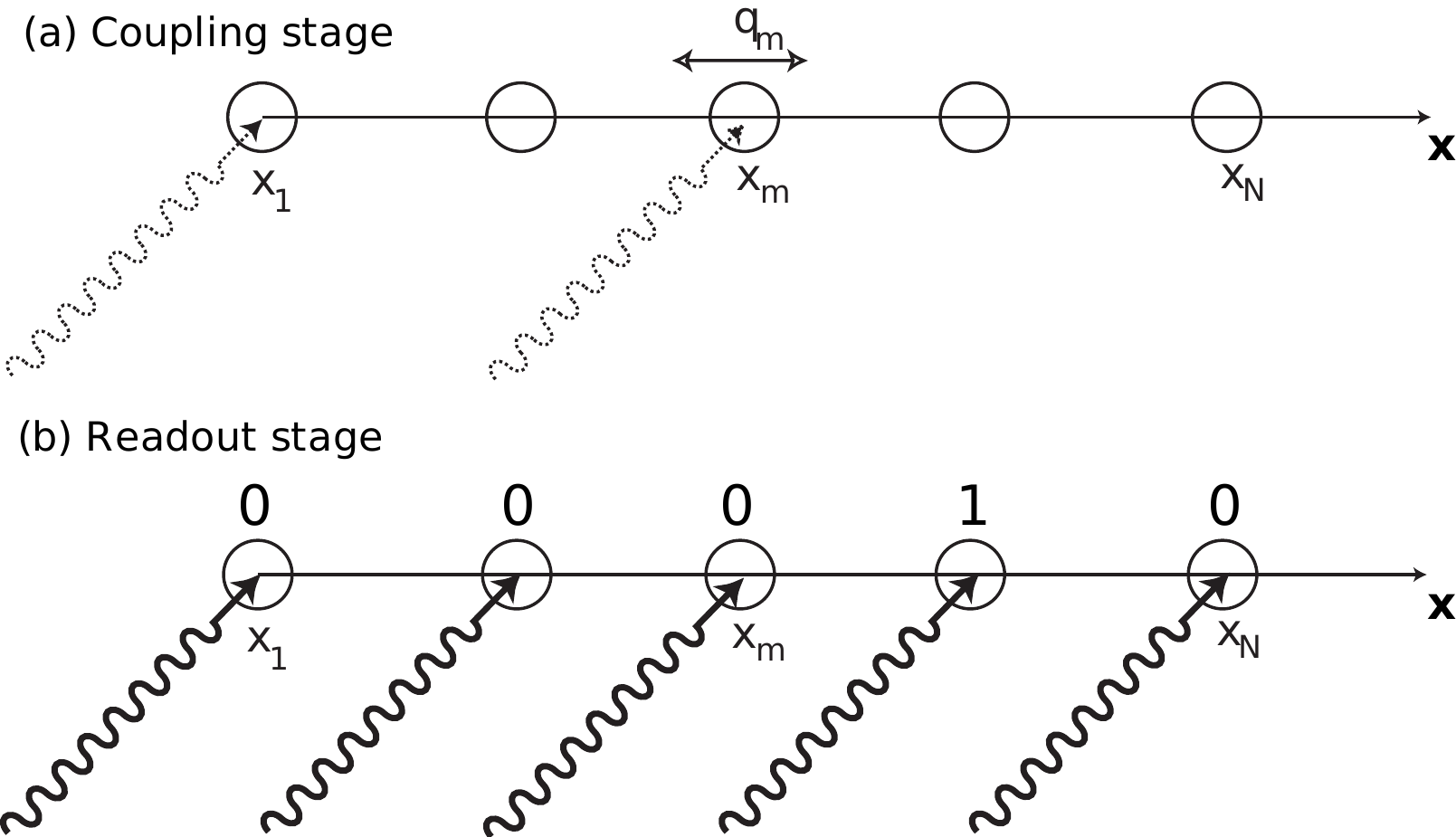}
\caption{An illustration of the linear ion array with $N$ ions. The index~$m$ runs from~1 to~$N$. This may be converted to a position label $x_m$ as indicated, which labels the equilibrium position of each ion. The quantum degree of freedom, $\hat{q}_m$ describes small displacements from equilibrium.  In~(a) weak lasers couple the internal electronic state of the ion to the displacement from equilibrium of that ion. In~(b)  strong readout lasers drive fluorescence conditional on the electronic state of the ion. If an a given ion is in the excited state, it fluoresces (giving the result~$1$) or it does not fluoresce (giving the result~$0$).} 
\label{fig:laser}
\end{figure}

These projectors commute for all~$m$, so we can represent joint detection of the excitation of multiple ions as
\begin{align}
\label{eq:calPmultiple}
	\mathcal P_{m_1 \dotsm m_M} &\coloneqq \mathcal P_{m_1} \dotsm \mathcal P_{m_M}\;.
\end{align}
We will be interested here in simultaneous detection on at most two ions,
\begin{align}
\label{eq:calPmn}
	\mathcal P_{mn} = \moutprod e e \otimes \noutprod e e
\end{align}
(assuming $m \neq n$, since $\mathcal P_{mm}$ is just $\mathcal P_m$), and initial states (in the interaction picture, at time~$t=0$) of the form
\begin{align}
\label{eq:rhogeneral}
	\rho_0 = \rho \otimes \outprod g g ^{\otimes N}\;,
\end{align}
where $\rho$ is the vibrational state, and all of the ions are in the ground electronic state.   The ions are assumed to be ordered in a linear array. Thus the subscripts $mn$ are an implicit spatial index. This is indicated schematically in Figure~\ref{fig:laser}.
%------------------
%------------------
The probability for measuring ion~$m$ in an excited electronic state after the interaction Hamiltonian is applied from time~$t=0$ to time~$t=T$ is given by
\begin{align}
\label{eq:Pm}
	P_m \coloneqq \langle U^\dag(T) \mathcal P_m U(T) \rangle\;,
\end{align}
where the expectation value is taken with respect to $\rho_0$.  Similarly, the probability that {\it both} ions~$m$ and~$n$ are in the excited electronic state after the evolution is
\begin{align}
\label{eq:Pmn}
	P_{mn} \coloneqq \langle U^\dag(T) \mathcal P_{mn} U(T) \rangle\;.
\end{align}
Since the only outcomes of an actual measurement is a stochastic binary string indicating which ions lit up and which did not~$(e_1, \dotsc, e_N)$, these probabilities correspond directly to classical expectation values over the entries~$e_m$ in this string:
\begin{align}
	\mathcal E[e_m] &= P_m\;, \nonumber \\
	\mathcal E[e_m e_n] &= P_{mn}\;,
\end{align}
Of course, other correlation functions (for three or more ions) may be defined analogously.

%%%%%%%%%%%%%%%%%%%%%%%%%%%%%%%%%%
% Excitation Probability Calculations for General States
%%%%%%%%%%%%%%%%%%%%%%%%%%%%%%%%%%

\section{Excitation Probability Calculations for General States}
\label{sec:generalstates}

We proceed to calculate $P_m$ and $P_{mn}$ to second order in the Dyson series expansion, Eq.~\eqref{eq:Uexpand}. The coupling lasers are assumed to be weak enough to justify this perturbative treatment.   The zeroth-order approximation $P_m^{(0)}$ vanishes since the initial state has no ions excited.  The first-order term is
\begin{align}
	P_m^{(1)} &= \frac {1} {\hbar^2} \inttzT dt_1 \inttzT dt_2 \, \langle H_I(t_1) \mathcal P_m H_I(t_2) \rangle_{\text{total}} \nonumber \\
	&= (\Omega_0 \eta)^2 \inttzT dt_1 \inttzT dt_2 \, \mbra g \hat \sigma_x^{(m)}(t_1) \moutprod e e \hat \sigma_x^{(m)}(t_2) \mket g \, \langle \hat \phi_m(t_1) \hat \phi_m(t_2) \rangle \;,
\end{align}
resulting in
% BOXED
\begin{align}
\label{eq:Pm1calc}
\boxed{
	P_m^{(1)} = (\Omega_0 \eta)^2 \inttzT dt_1 \inttzT dt_2 \, e^{-i\Delta(t_1 - t_2)} \langle \hat \phi_m (t_1) \hat \phi_m(t_2) \rangle \;,
	}
\end{align}
where expectation values are now over the vibrational modes only unless otherwise indicated.  The second-order term $P_m^{(2)} = 0$ because $\mbra g H_I(t_1) H_I(t_2) \mket e \sim \mbra g \sigma_\pm^{(m)} \sigma_\pm^{(m)} \mket e = 0$---i.e., no {\it two} applications of the electronic raising/lowering operators can lower the excited state to the ground state.

We're now prepared to tackle the spatial correlation function~\eqref{eq:Pmn}.  Once again, the zeroth-order term $P_{mn}^{(0)} = 0$ because no ions are electronically excited to begin with.  Also, considering that $\mathcal P_{mn}$ projects onto the excited electronic states of two distinct ions while each $H_I(t)$ can only raise one ion at a time, $\mbra g \otimes \nbra g H_I(t_1) \mket e \otimes \nket e = 0$, making $P_{mn}^{(1)} = 0$, as well.  Therefore, we must go to second order:
\begin{align}
	P_{mn}^{(2)} &= \frac {1}{4\hbar^4} \inttzT dt_1 \inttzT dt_2 \inttzT dt_3 \inttzT dt_4 \, \langle \bar \timeord \{ H_I(t_2) H_I(t_1) \} \mathcal P_{mn} \timeord \bigl\{ H_I(t_3) H_I(t_4) \bigl\} \rangle_{\text{total}} \nonumber \\
	&= \frac {1}{4\hbar^4} \inttzT d^4 t \, \langle \bar \timeord \bigl\{ [ H_I^{(m)}(t_2) + H_I^{(n)}(t_2) ] [ H_I^{(m)}(t_1) + H_I^{(n)}(t_1) ] \bigl\} \mathcal P_{mn} \nonumber \\
	&\hphantom{=} \hphantom{\frac {1}{4\hbar^4} \inttzT d^4 t \,} \qquad \times \timeord \bigl\{ [ H_I^{(m)}(t_3) + H_I^{(n)}(t_3) ] [ H_I^{(m)}(t_4) + H_I^{(n)}(t_4) ] \bigl\} \rangle_{\text{total}} \nonumber \\
	&= \frac {1}{4\hbar^4} \inttzT d^4 t \, \langle \bar \timeord \{ H_I^{(m)}(t_2) H_I^{(n)}(t_1) + H_I^{(n)}(t_2) H_I^{(m)}(t_1) \}  \mathcal P_{mn} \nonumber \\
	&\hphantom{=} \hphantom{\frac {1}{4\hbar^4} \inttzT d^4 t \,} \qquad \times \timeord \{ H_I^{(m)}(t_3) H_I^{(n)}(t_4) + H_I^{(n)}(t_3) H_I^{(m)}(t_4) \} \rangle_{\text{total}}\;.
\end{align}
The antitime-ordering symbol~$\bar \timeord$ acts like~$\timeord$ but instead orders the terms from earliest to latest.  Trading integration variables ($t_1 \Leftrightarrow t_2$ and/or $t_3 \Leftrightarrow t_4$) and minding the (anti)time-ordering, we can simplify this to
\begin{align}
	P_{mn}^{(2)} &= \frac {1}{\hbar^4} \inttzT d^4 t \langle \bar \timeord \{ H_I^{(m)}(t_2) H_I^{(n)}(t_1) \}  \mathcal P_{mn} \timeord \{ H_I^{(m)}(t_3) H_I^{(n)}(t_4) \} \rangle_{\text{total}}\;.
\end{align}
The terms from the Hamiltonians that survive the projection and contraction with the electronic ground state are
\begin{align}
\label{eq:Hsurviveleft}
	\bar \timeord \{H_I^{(m)}(t_2) H_I^{(n)} (t_1) \} &\Longrightarrow (\hbar \Omega_0 \eta)^2 \bar \timeord\{ \hat \phi_m(t_2) \hat \phi_n(t_1) \} e^{-i\Delta t_2} e^{-i\Delta t_1} \sigma_-^{(m)} \sigma_-^{(n)} \\
\label{eq:Hsurviveright}
	\text{and}\qquad \timeord \{ H_I^{(m)}(t_3) H_I^{(n)} (t_4) \} &\Longrightarrow (\hbar \Omega_0 \eta)^2 \timeord \{ \hat \phi_m(t_3) \hat \phi_n(t_4) \} e^{+i\Delta t_3} e^{+i\Delta t_4} \sigma_+^{(m)} \sigma_+^{(n)}\;.
\end{align}
Plugging these in gives
% BOXED
\begin{align}
\label{eq:Pmn2calc}
\boxed{
	P_{mn}^{(2)} = (\Omega_0 \eta)^4 \inttzT d^4 t\, e^{-i\Delta(t_1 + t_2 - t_3 - t_4)} \langle \bar \timeord \{ \hat \phi_m(t_2) \hat \phi_n(t_1) \} \timeord \{ \hat \phi_m(t_3) \hat \phi_n(t_4) \} \rangle\;.
	}
\end{align}
Several ways exist for expressing the four-point correlation function in Eq.~\eqref{eq:Pmn2calc} in a convenient form by use of Wick's theorem.  These are included in the~Appendix.  One form is particularly useful, though, and that it applies when the state has a Wigner function that is a Gaussian with zero mean (i.e., a ``Gaussian state'').  In this case, we may write (repeated from Eq.~\eqref{eq:Pmn3terms})
\begin{multline}
\label{eq:Pmn3termsbody}
	\langle \bar \timeord \{ \hat \phi_m(t_2) \hat \phi_n(t_1) \} \timeord \{ \hat \phi_m(t_3) \hat \phi_n(t_4) \} \rangle \xrightarrow[\text{state}]{\text{Gaussian}} \langle \hat \phi_m(t_2) \hat \phi_m(t_3) \rangle \langle \hat \phi_n(t_1) \hat \phi_n(t_4) \rangle \\
	+ \langle \hat \phi_m(t_2) \hat \phi_n(t_4) \rangle \langle \hat \phi_n(t_1) \hat \phi_m(t_3) \rangle 
	+ \langle \bar \timeord \{ \hat \phi_m(t_2) \hat \phi_n(t_1) \} \rangle \langle \timeord \{ \hat \phi_m(t_3) \hat \phi_n(t_4) \} \rangle\;.
\end{multline}
Comparing this with Eq.~\eqref{eq:Pm1calc}, we see that the first term will always give simply $P_m P_n$, and the second will always give a term similar to this, but with a different geometric factor.  Thus, once we have~$P_m$, we need only ever explicitly calculate the last two terms from Eq.~\eqref{eq:Pmn3termsbody} to get~$P_{mn}$.

%
% Long-time interaction
%

\subsection{Long-time interaction}
\label{subsec:RWA}

Before moving on, let's have a look at what affect a long-time interaction would have on the evolution.  In this case, we can employ the rotating wave approximation and use Eq.~\eqref{eq:HIredmodep} as the interaction Hamiltonian.  In this case, Eq.~\eqref{eq:Udef} can be evaluated to
\begin{align}
\label{eq:UinRWA}
	U(T) &\xrightarrow[\text{($\Delta = \nu_p$)}]{\text{RWA}} \exp \left[-i T (\Omega_0 \eta)  {\sum_m}' \frac {b_m^{(p)}} {\mu_p^{1/4}} (a_p \sigma_+^{(m)} + a_p^\dag \sigma_-^{(m)}) \right]\;.
\end{align}
There is a balance to be maintained, here, though.  On the one hand, the detection time~$T$ needs to be long enough so that the rotating wave approximation is valid.  This is the requirement that~$T \gg \nu^{-1}$ (and we assume that $\Delta \sim \nu$).  On the other hand, it needs to be short enough so that we can use perturbation theory, which requires~$T \ll (\Omega_0 \eta)^{-1}$.  When both of these conditions are satisfied simultaneously, that is,
\begin{align}
\label{eq:RWAcond}
	\nu^{-1} \ll T \ll (\Omega_0 \eta)^{-1}\;,
\end{align}
we can expand the exponential in Eq.~\eqref{eq:UinRWA}, as in Eq.~\eqref{eq:Uexpand}.  This gives
\begin{align}
\label{eq:Pm1RWA}
	P_m^{(1)} &\xrightarrow[\text{($\Delta = \nu_p$)}]{\text{RWA}} \frac {T^2} {\hbar^2} \avg{ H_I \mathcal P_{m} H_I }_\text{total} = T^2 (\Omega_0 \eta)^2 \frac {b_m^{(p)2}} {\sqrt{\mu_p}} \avg{ a_p^\dag a_p }\;,
\end{align}
which agrees with what is obtained from selecting only the resonant terms directly from Eq.~\eqref{eq:Pm1calc}.  A similar result holds for the second-order function:
\begin{align}
\label{eq:Pmn2RWA}
	P_{mn}^{(2)} &\xrightarrow[\text{($\Delta = \nu_p$)}]{\text{RWA}} \frac {T^4} {4 \hbar^4} \avg{ H_I^2 \mathcal P_{mn} H_I^2 }_{\text{total}} = T^4 (\Omega_0 \eta)^4 \frac {(b_m^{(p)} b_n^{(p)})^2} {\mu_p} \avg{a_p^\dag a_p^\dag a_p a_p}\;.
\end{align}
If we were to allow for different detunings on each ion, $\Delta_m = \nu_p$, $\Delta_n = \nu_{p'}$, then this result could be generalized:
\begin{align}
\label{eq:Pmn2RWAdetune}
	P_{mn}^{(2)} &\xrightarrow[\text{($\Delta_m = \nu_p$, $\Delta_n = \nu_{p'}$)}]{\text{RWA}} \frac {T^4} {4 \hbar^4} \avg{ H_I^2 \mathcal P_{mn} H_I^2 }_{\text{total}} = T^4 (\Omega_0 \eta)^4 \frac {(b_m^{(p)} b_n^{(p')})^2} {\sqrt{ \mu_p \mu_{p'}}} \avg{a_p^\dag a_{p'}^\dag a_{p'} a_p}\;.
\end{align}
This interaction would require that a second laser beam be detuned from the first, an experimental complication we won't consider further.

For typical values of the parameters ($\nu \sim 1~\text{MHz}$, $\Omega_0 \sim 100~\text{kHz}$, and $\nu \sim 0.01$), Condition~\eqref{eq:RWAcond} requires that $10^{-6}~\text{s} \ll T \ll 10^{-3}~\text{s}$, a rather narrow band in which to operate well within both regimes.  Given this limitation, in the next section, when we calculate~$P_m$ and~$P_{mn}$ for any Gaussian state as a function of its covariance matrix, we will retain the perturbative condition~$T \ll (\Omega_0 \eta)^{-1}$, but we will not use the rotating wave approximation, in order to allow for a more general class of interactions, including ones for which $T \lesssim \nu^{-1}$.  We will also assume that both interaction lasers have the same detuning~$\Delta$ for experimental simplicity.

%%%%%%%%%%%%%%%%%%%%%%%%%%%%%%%%%%
% Evaluation in Terms of the Covariance Matrix
%%%%%%%%%%%%%%%%%%%%%%%%%%%%%%%%%%

\section{Evaluation for Gaussian States}
\label{sec:evalgaussian}

\subsection{Two-point functions}

As discussed in the Appendix, when the vibrational state is a zero-mean Gaussian, all measured correlation functions can be evaluated from the two-point functions~$\avg{\phi_m(t) \phi_n(t')}$ and $\avg{ \timeord \{ \phi_m(t) \phi_n(t') \} }$.  Let's define the two-time-dependent matrices
\begin{align}
\label{eq:Upsilondef}
	\mat \Upsilon(t,t') &\coloneqq \avg{ \vec \phi(t) \vec \phi(t')^T } \\
\label{eq:barUpsilondef}
	\text{and} \qquad \mathring {\mat \Upsilon}(t,t') &\coloneqq \avg{ \timeord \{ \vec \phi(t) \vec \phi(t')^T \} } \\
\end{align}
for which $\Upsilon_{mn}(t,t') = \avg{\phi_m(t) \phi_n(t')}$ and $\mathring \Upsilon_{mn}(t,t') = \avg{ \timeord \{ \phi_m(t) \phi_n(t') \} }$.  It's easy to see from the entries that
\begin{align}
\label{eq:ringUpsilon}
	\mathring {\mat \Upsilon}(t,t') =
	\begin{cases}
		\mat \Upsilon(t,t') & \text{if $t > t'$}\;, \\
		\mat \Upsilon(t,t')^* & \text{if $t < t'$}\;,
	\end{cases}
\end{align}
since $\phi_m(t)$~is Hermitian.  Thus, we really only need to worry about~$\mat \Upsilon(t,t')$ for the moment.  Using Eqs.~\eqref{eq:phidef} and~\eqref{eq:Adiag}, we can transform to the normal-mode two point function instead:
\begin{align}
\label{eq:Upsilonnorm}
	\mat \Upsilon(t,t') &= \mat B^T \mat \Lambda^{-1/4} \avg{ \bigl[ \vec a(t) + \tilde {\vec a}(t) \bigr] \bigl[ \vec a(t') + \tilde {\vec a}(t') \bigr]^T } \mat \Lambda^{-1/4} \mat B\;,
\end{align}
where $\vec a(t) \coloneqq (a_1(t), \dotsc, a_N(t))^T$~is a column vector of interaction-picture lowering operators for the normal modes, $\tilde {\vec a}(t) \coloneqq (a^\dag_1(t), \dotsc, a^\dag_N(t))^T$~is the equivalent column vector of interaction-picture raising operators, and~$\mat B$ and~$\mat \Lambda$ are defined through Eq.~\eqref{eq:Adiag}.  We can write the time-dependence of these vectors in a very compact form:
\begin{align}
	\vec a(t) &= \vec E(t) \hadam \vec a\;,
\end{align}
where
\begin{align}
\label{eq:vecEoft}
	\vec E(t) &\coloneqq (e^{-i \nu_1 t}, \dotsc, e^{-i \nu_N t})^T
\end{align}
is a column vector of time-dependent coefficients, and the symbol~$\hadam$ represents the Hadamard product (element-wise multiplication).  Similarly,
\begin{align}
	\tilde {\vec a}(t) &= \vec E(-t) \hadam \tilde{\vec a}\;.
\end{align}
Any Gaussian state with zero mean is uniquely defined by its covariance matrix.  While many varieties of covariance matrix can be defined~\cite{Walls2008}, an obvious choice here would be to use the two matrices~$\avg{\vec a \vec a^T}$ and~$\avg{\tilde{\vec a} \vec a^T}$.  We can get the other combinations by noting that~$\avg{\tilde{\vec a} \tilde{\vec a}^T} = \avg{\vec a \vec a^T}^*$, and~$\avg{\vec a \tilde{\vec a}^T} = \avg{\tilde{\vec a} \vec a^T} + \1$.  Thus, we can go one step further and define a matrix of coefficients
\begin{align}
	\mat E(t, t') \coloneqq \vec E(t) \vec E(t')^T\;,
\end{align}
such that $E_{rs}(t,t') = e^{-i(\nu_r t + \nu_s t')}$.  Using this shorthand, we can isolate the time dependence from the expectation value in Eq.~\eqref{eq:Upsilonnorm} and write the result in terms of the initial covariance matrix:
\begin{align}
\label{eq:Kdef}
	\mat K(t,t') &:= \avg{ \bigl[ \vec a(t) + \tilde {\vec a}(t) \bigr] \bigl[ \vec a(t') + \tilde {\vec a}(t') \bigr]^T } \nonumber \\
	& = \avg{\vec a \vec a^T} \hadam \mat E(t,t')
	+	\avg{\vec a \tilde{\vec a}^T} \hadam \mat E(t,-t')
	+	\avg{\tilde{\vec a} \vec a^T} \hadam \mat E(-t, t')
	+	\avg{\tilde{\vec a} \tilde{\vec a}^T} \hadam \mat E(-t,-t') \nonumber \\
	& = 2 \Realpart \left[ \avg{\vec a \vec a^T} \hadam \mat E(t,t') \right]
	+	\left[\avg{\tilde{\vec a} \vec a^T} + \1\right] \hadam \mat E(t,-t')
	+	\avg{\tilde{\vec a} \vec a^T} \hadam \mat E(-t, t')\;.
\end{align}

Considering Eq.~\eqref{eq:ringUpsilon}, we're going to need the complex conjugate of Eq.~\eqref{eq:Kdef}.  Element-by-element evaluation will reveal the following equivalences:
\begin{align}
	\left[ \avg{\vec a \tilde{\vec a}^T} \hadam \mat E(t,-t') \right]^* &= \left[ \avg{\tilde{\vec a} \vec a^T} + \1 \right] \hadam \mat E(-t, t')\;, \\
	\left[ \avg{\tilde{\vec a} \vec a^T} \hadam \mat E(-t, t')\right]^* &= \left[ \avg{\vec a \tilde{\vec a}^T} - \1 \right] \hadam \mat E(t,-t')\;,
\end{align}
leading to
\begin{align}
	&\left[ \avg{\vec a \tilde{\vec a}^T} \hadam \mat E(t,-t') + \avg{\tilde{\vec a} \vec a^T} \hadam \mat E(-t, t')\right]^* \nonumber \\
	&\qquad = \avg{\vec a \tilde{\vec a}^T} \hadam \mat E(t,-t') + \avg{\tilde{\vec a} \vec a^T} \hadam \mat E(-t, t') + \1 \hadam \left[ \mat E(-t,t') - \mat E(t, -t') \right] \nonumber \\
	&\qquad = \left[ \avg{\tilde{\vec a} \vec a^T} + \1 \right]\hadam \mat E(-t, t') + \avg{\tilde{\vec a} \vec a^T} \hadam \mat E(t,-t')\;.
\end{align}
This is the same as the last two terms in Eq.~\eqref{eq:Kdef}, except that the $\mat E$-matrices have been exchanged.  Therefore, we can define
\begin{align}
	\mathring{\mat E}(t,t') :=
	\begin{cases}
		\mat E(t,-t') & \text{if $t > t'$}\;, \\
		\mat E(-t,t') & \text{if $t < t'$}\;.
	\end{cases}
\end{align}
Notice the subtle difference between~$\mathring{\mat \Upsilon}(t,t')$ and~$\mathring{\mat E}(t,t')$.  The ring-notation is consistent with its purpose---to describe a function associated with time-ordering---but not necessarily with the details of the definition.  Recalling Eq.~\eqref{eq:ringUpsilon}, the time-ordered expectation value can now be written succinctly:
\begin{align}
\label{eq:ringKdef}
	\mathring{\mat K}(t,t') &:= \avg{ \timeord \left\{ \bigl[ \vec a(t) + \tilde {\vec a}(t) \bigr] \bigl[ \vec a(t') + \tilde {\vec a}(t') \bigr]^T \right\} } \nonumber \\
	\qquad &= 2 \Realpart \left[ \avg{\vec a \vec a^T} \hadam \mat E(t,t') \right]
	+	\left[\avg{\tilde{\vec a} \vec a^T} + \1\right] \hadam \mathring{\mat E}(t,t')
	+	\avg{\tilde{\vec a} \vec a^T} \hadam \mathring{\mat E}(t, t')^*\;.
\end{align}
Pluggin into Eqs.~\eqref{eq:Upsilonnorm} and~\eqref{eq:ringUpsilon} gives
\begin{align}
	\mat \Upsilon(t,t') &= \mat B^T \mat \Lambda^{-1/4} \mat K(t,t') \mat \Lambda^{-1/4} \mat B\;, \\
	\mathring{\mat \Upsilon}(t,t') &= \mat B^T \mat \Lambda^{-1/4} \mathring{\mat K}(t,t') \mat \Lambda^{-1/4} \mat B\;,
\end{align}
with $\mat K(t,t')$~defined in Eq.~\eqref{eq:Kdef} and $\mathring{\mat K}(t,t')$~in Eq.~\eqref{eq:ringKdef}.

%
% Probabilities in terms of the covariance matrix
%

\subsection{Probabilities in terms of the covariance matrix}

We have successfully consolidated all of the time-dependence into~$\mat E(t,t')$ and~$\mathring{\mat E}(t,t')$.  This makes evaluation of Eqs.~\eqref{eq:Pm1calc} and~\eqref{eq:Pmn2calc} dependent only on integrals involving elements of these matrices.  The following integral will be useful:
\begin{align}
\label{eq:sincint}
	\frac 1 T \intsym dt\, e^{i \omega t} =  \sinc \left[ \frac {\omega T} {2} \right]\;,
\end{align}
where
\begin{align}
\label{eq:sincdef}
	\sinc x \coloneqq
	\begin{cases}
		x^{-1} \sin x & \text{if $x\ne 0$}\;, \\
		1 & \text{if $x=0$}\;,
	\end{cases}
\end{align}
and we have symmetrized the limits of integration by setting the laboratory clock appropriately---a passive operation that does not affect the physics of the experiment.  Eq.~\eqref{eq:sincint} is a bandwidth-limited Fourier transform.  In anticipation of future calculations, let's define
\begin{align}
	S_{rs}(\omega,\omega') &\coloneqq \frac {1} {T^2} \intsym dt \intsym dt' \, e^{i (\omega t + \omega' t')} E_{rs}(t,t') \nonumber \\
	&= \sinc \left[ \frac {(\omega - \nu_r) T} {2} \right] \sinc \left[ \frac {(\omega' - \nu_s) T} {2} \right]\;.
\end{align}
We can now collect the~$S_{rs}(\omega,\omega')$ elements into a matrix~$\mat S(\omega, \omega')$.  We can also define
\begin{align}
\label{eq:ringSdef}
	\mathring S_{rs}(\omega,\omega') &\coloneqq \frac {1} {T^2} \intsym dt \intsym dt' \, e^{i (\omega t + \omega' t')} \mathring E_{rs}(t,t')\;,
\end{align}
and a corresponding matrix~$\mathring{\mat S}(\omega,\omega')$, although we will leave it unevaluated for now.

Using these tools, let's calculate~$P_m^{(1)}$:
\begin{align}
\label{eq:Pm1Gauss}
	P_m^{(1)} &= (\Omega_0 \eta)^2 \intsym dt_1 \intsym dt_2 \, e^{-i\Delta(t_1 - t_2)} \Upsilon_{mm}(t_1,t_2) \nonumber \\
	&= (\Omega_0 \eta)^2 \vec e_m^T \mat B^T \mat \Lambda^{-1/4}
		\left[\intsym dt_1 \intsym dt_2 \, e^{-i\Delta(t_1 - t_2)} \mat K(t_1,t_2)\right]
		\mat \Lambda^{-1/4} \mat B \vec e_m \nonumber \\
	&= T^2 (\Omega_0 \eta)^2 \vec e_m^T \mat B^T \mat \Lambda^{-1/4} \biggl[
		\avg{\vec a \vec a^T} \hadam \mat S(-\Delta,\Delta)
	+	\avg{\tilde{\vec a} \tilde{\vec a}^T} \hadam \mat S(\Delta,-\Delta) \nonumber \\
	&\qquad \qquad \qquad
	+	\left[\avg{\tilde{\vec a} \vec a^T} + \1\right] \hadam \mat S(-\Delta,-\Delta)
	+	\avg{\tilde{\vec a} \vec a^T} \hadam \mat S(\Delta,\Delta)
	\biggr] \mat \Lambda^{-1/4} \mat B \vec e_m\;,
\end{align}
where $\vec e_m$~is a unit column vector used (twice) to pick out the correct element from the matrix.  We should also point out that Eq.~\eqref{eq:Pm1Gauss} also holds for non-Gaussian states, for which a covariance matrix can also be defined even though it does not specify the state completely.

The correlation function~$P_{mn}$ can be calculated similarly.  In this case, \emph{it is} important that the state be (zero-mean) Gaussian because we will use the simplification provided by Eq.~\eqref{eq:Pmn3termsbody}.  As already discussed,
\begin{align}
\label{eq:term1Gauss}
	\text{(Term~1)} = P_m P_n\;.
\end{align}
The second is almost the same, except it involves a different element of~$\mat \Upsilon(t,t')$.  An analogous calculation to the one above shows that
\begin{multline}
\label{eq:term2Gauss}
	\text{(Term~2)} = T^4 (\Omega_0 \eta)^4 \biggl\lvert \vec e_m^T \mat B^T \mat \Lambda^{-1/4} \biggl[
		\avg{\vec a \vec a^T} \hadam \mat S(-\Delta,\Delta)
	+	\avg{\tilde{\vec a} \tilde{\vec a}^T} \hadam \mat S(\Delta,-\Delta) 
		\\
	+	\left[\avg{\tilde{\vec a} \vec a^T} + \1\right] \hadam \mat S(-\Delta,-\Delta)
	+	\avg{\tilde{\vec a} \vec a^T} \hadam \mat S(\Delta,\Delta)
	\biggr] \mat \Lambda^{-1/4} \mat B \vec e_n \biggr\rvert^2\;,
\end{multline}
the only difference (besides the squaring) being the presence of~$\vec e_n$ at the end, instead of~$\vec e_m$.  The third term is more complicated, due to the time ordering.  Nevertheless, it can be written
\begin{align}
\label{eq:term3Gauss}
	\text{(Term~3)} &= (\Omega_0 \eta)^4 \abs{ \intsym dt_1 \intsym dt_2 \, e^{i\Delta(t_1 + t_2)} \mathring \Upsilon_{mn}(t_1,t_2) }^2 \nonumber \\
	&= (\Omega_0 \eta)^4 \abs{ \vec e_m^T \mat B^T \mat \Lambda^{-1/4}
		\left[\intsym dt_1 \intsym dt_2 \, e^{i\Delta(t_1 + t_2)} \mathring{\mat K}(t_1,t_2)\right]
		\mat \Lambda^{-1/4} \mat B \vec e_n }^2 \nonumber \\
	&= T^4 (\Omega_0 \eta)^4 \left\lvert \vec e_m^T \mat B^T \mat \Lambda^{-1/4} \biggl[
		\avg{\vec a \vec a^T} \hadam \mat S(\Delta,\Delta)
	+	\avg{\tilde{\vec a} \tilde{\vec a}^T} \hadam \mat S(-\Delta,-\Delta) \right. \nonumber \\
	&\qquad \qquad \left.
	+	\left[\avg{\tilde{\vec a} \vec a^T} + \1\right] \hadam \mathring{\mat S}(\Delta,\Delta)
	+	\avg{\tilde{\vec a} \vec a^T} \hadam \mathring{\mat S}(-\Delta,-\Delta)^*
	\biggr] \mat \Lambda^{-1/4} \mat B \vec e_n \right\rvert^2\;.
\end{align}
Evaluating this term boils down to evaluating Eq.~\eqref{eq:ringSdef}.  The sum of Terms~1, 2, and~3 gives~$P_{mn}^{(2)}$ for a zero-mean Gaussian state.

%%%%%%%%%%%%%%%%%%%%%%%%%%%%%%%%%%
% Examples
%%%%%%%%%%%%%%%%%%%%%%%%%%%%%%%%%%

\section{Examples}
\label{sec:examples}

In order to demonstrate the usefulness of the correlation function, we'll compare a thermal state with a given average phonon number to a corresponding uniformly squeezed state with the same average phonon number.  We will also assume that the interaction is active on a timescale long compared to the period of vibrations of the ions---that is, $T \gg \nu^{-1}$.  In both cases, the probability of excitation for any given ion is the same, but, as we shall see, the correlations are stronger in the squeezed state versus the corresponding thermal state.  Our measure of correlations will be
\begin{align}
	f_{mn} \coloneqq \frac {P_{mn}} {P_m P_n} = \frac {\mathcal E[e_m e_n]} {\mathcal E[e_m] \mathcal E[e_n]}\;,
\end{align}
which is a normalized correlation function akin to those defined in Eqs.~\eqref{eq:normcorr} but is based on detection probabilities, rather than underlying quantum correlations.  The connection to the quantum state, of course, was made in the preceding sections.

%
% Thermal State
%

\subsection{Thermal state}
\label{subsec:thermal}

A thermal state at temperature~$\tau$ is a zero-mean Gaussian state in which
\begin{align}
	\avg{a_r a_s} &= 0\;, \\
\label{eq:adagathermal}
	\avg{a^\dag_r a_s} &= \bar n_r \delta_{rs}\;,
\end{align}
where
% $n_r \coloneqq (e^{\beta \hbar \nu_r} - 1)^{-1}$ 
\begin{align}
\label{eq:barnr}
	n_r \coloneqq \frac {1} {e^{\beta \hbar \nu_r} - 1}
\end{align}
is the average number of phonons in mode~$r$, and $\beta = (k_\boltz \tau)^{-1}$.  Plugging into Eq.~\eqref{eq:Pm1Gauss} gives
\begin{align}
\label{eq:Pm1thermal}
%\boxed{
	P_m^{(1)} = T^2 (\Omega_0 \eta)^2 \sum_{p=1}^N \frac {b_m^{(p)2}} {\sqrt {\mu_p}} \Bigl\{ (\bar n_p + 1)  \sinc^2 \bigl[ (\Delta + \nu_p) \tfrac T 2 \bigr] + \bar n_p \sinc^2 \bigl[ (\Delta - \nu_p) \tfrac T 2 \bigr] \Bigr\}\;.
%	}
\end{align}
The $\sinc^2$-function is sharply peaked at $\Delta = \pm \nu_p$, and in the long-time limit ($T \gg \nu^{-1}$), we have
\begin{align}
\label{eq:Pm1thermallimit}
	P_m^{(1)} \xrightarrow{\text{($T \gg \nu^{-1}$)}} T^2 (\Omega_0 \eta)^2 \sum_p \frac {b_m^{(p)2}} {\sqrt {\mu_p}} \Bigl\{ (\bar n_p + 1) (2\pi)^2 \delta_{(\Delta + \nu_p)} + \bar n_p (2\pi)^2 \delta_{(\Delta - \nu_p)} \Bigr\}\;,
\end{align}
where the Kronecker-$\delta$ symbol satisfies $\delta_0 = 1$ and $\delta_x = 0$ for~$x \ne 0$ to within a bandwidth of approximately~$T^{-1}$.  This is gives the same results as would be obtained after using the rotating wave approximation, as described in Section~\ref{subsec:RWA}.  Choosing a particular normal mode frequency as the detuning~($\Delta = \nu_p$) accords with the result calculated directly from Eq.~\eqref{eq:Pm1RWA}.  In this case, the detection probability has a geometric factor~$\mu_p^{-1/2} b_m^{(p)}$, corresponding to the position of the ion within the normal mode~$p$ being addressed and is proportional to the average number of phonons~$\bar n_p$ in the mode for red sideband detuning (and $\bar n_p+1$ for the blue sideband), as expected.

Moving on to the correlation probability~$P_{mn}^{(2)}$, we can evaluate Eq.~\eqref{eq:term2Gauss} to
\begin{multline}
\label{eq:Pmn2term2thermal}
	\text{(Term~2)} = \\
	T^4 (\Omega_0 \eta)^4 \left[ \sum_a \frac {b_m^{(p)} b_n^{(p)}} {\sqrt {\mu_p}} \Bigl\{ (\bar n_p + 1)  \sinc^2 \bigl[ (\Delta + \nu_p) \tfrac T 2 \bigr] + \bar n_p \sinc^2 \bigl[ (\Delta - \nu_p) \tfrac T 2 \bigr] \Bigr\} \right]^2\;.
\end{multline}
In the long interaction time-limit, this term behaves similarly to~$P_m^{(1)}$:
\begin{multline}
\label{eq:Pmn2term2thermallimit}
	\text{(Term~2)}
	 \xrightarrow{\text{($T \gg \nu^{-1}$)}} T^4 (\Omega_0 \eta)^4 \left[ \sum_a \frac {b_m^{(p)} b_n^{(p)}} {\sqrt {\mu_p}} \Bigl\{ (\bar n_p + 1) (2\pi)^2 \delta_{(\Delta + \nu_p)} + \bar n_p (2\pi)^2 \delta_{(\Delta - \nu_p)} \Bigr\} \right]^2 \;.
\end{multline}

In order to evaluate Eq.~\eqref{eq:term3Gauss}, we need to evaluate $\mathring{\mat S}(\pm \Delta, \pm \Delta)$ from Eq.~\eqref{eq:ringSdef}.  Using Eq.~\eqref{eq:adagathermal}, $\mathring {\mat E}(t_1,t_2)$~is diagonal, meaning we only need to calculate
\begin{align}
	\mathring S_{pp}(\pm \Delta,\pm \Delta) &= \frac {1} {T^2} \intsym dt_1 \intsym dt_2 \, e^{\pm i \Delta( t_1 + t_2)} \mathring E_{pp}(t_1,t_2) \nonumber \\
	&= \frac {1} {T^2} \intsym dt_1 \intsym dt_2 \, e^{\pm i \Delta( t_1 + t_2)} e^{-i\nu_p \abs{t_1-t_2}} \nonumber \\
	&= \frac {2} {T^2} \intsym dt_1 \int_{-T/2}^t dt_2 \, e^{\pm i \Delta( t_1 + t_2)} e^{-i\nu_p (t_1-t_2)}\;.
\end{align}
Since $T \gg \nu^{-1} \sim \Delta$, we can change the integration limits with~$T \to \infty$:
\begin{align}
	\mathring S_{pp}(\pm \Delta,\pm \Delta) &= \frac {2} {T^2} \infint dt_1 \int_{-\infty}^t dt_2 \, e^{\pm i \Delta( t_1 + t_2)} e^{-i\nu_p (t_1-t_2)}\;.
\end{align}
With the integration now over the entire half-plane defined by $t_2 < t_1$, we can rotate our integration axes using
\begin{equation}
\label{eq:rotatedcoordsthermal}
\begin{aligned}
	t_1 &= u + v	&\qquad&& u &= \frac 1 2 (t_1 + t_2) \\
	t_2 &= u - v	&\qquad&& v &= \frac 1 2 (t_1 - t_2)
\end{aligned}
\end{equation}
to obtain
\begin{align}
\label{eq:I2Drotatedthermal}
	\mathring S_{pp}(\pm \Delta,\pm \Delta) &= 2 \halfint dv \infint du \, e^{\pm i2\Delta u} e^{-i2\nu_p v} \sim \delta_\Delta \delta_{\nu_p} \to 0\;.
\end{align}
Thus, in the thermal case,
\begin{align}
\label{eq:Pmn2term3thermal}
	\text{(Term~3)} \xrightarrow{\text{($T \gg \nu^{-1}$)}} 0\;.
\end{align}
Consolidating these results, we have
\begin{align}
\label{eq:fthermal}
	f_{mn} \simeq \frac {P_{mn}^{(2)}} {P_m^{(1)} P_n^{(1)}} = 2\;,
\end{align}
for any detuning~$\Delta = \pm \nu_p$, any nonzero temperature, and any choice of ions~$(m,n)$ to measure.  This is consistent with the expected results for a second-order correlation function for a thermal state~\cite{Walls2008}.

%
% Uniformly Squeezed Normal Modes
%

\subsection{Uniformly squeezed normal modes}
\label{subsec:squeezed}

The state to be considered in this section is uniformly squeezed in all normal modes.  Such a state has
 \begin{align}
	\avg{a_r a_s} &= -\sqrt{\bar n (\bar n + 1)}\,  \delta_{rs}\;, \\
\label{eq:adagasqueezed}
	\avg{a^\dag_r a_s} &= \bar n \, \delta_{rs}\;,
\end{align}
where $\bar n = \sinh^2 r$ is the mean phonon number of each mode (assumed the same for each mode) as a function of the squeezing parameter~$r$.

Proceeding as for the thermal state, we have nearly the same expression for $P_m^{(1)}$:
\begin{multline}
\label{eq:Pm1squeezed}
%\boxed{
	P_m^{(1)} = T^2 (\Omega_0 \eta)^2 \sum_{p=1}^N \frac {b_m^{(p)2}} {\sqrt {\mu_p}} \Bigl\{ (\bar n + 1)  \sinc^2 \bigl[ (\Delta + \nu_p) \tfrac T 2 \bigr] + \bar n \sinc^2 \bigl[ (\Delta - \nu_p) \tfrac T 2 \bigr]  \\
		- 2\sqrt{\bar n (\bar n + 1)} \sinc \bigl[ (\Delta - \nu_p) \tfrac T 2 \bigr]  \sinc \bigl[ (\Delta + \nu_p) \tfrac T 2 \bigr] \Bigr\}\;.
%	}
\end{multline}
In the long detection time-limit, however, the last term makes no difference, and
\begin{align}
	P_{m,\text{squeezed}}^{(1)} \xrightarrow{\text{($T \gg \nu^{-1}$)}} P_{m,\text{thermal}}^{(1)}\;,
\end{align}
where $P_{m,\text{thermal}}^{(1)}$~is Eq.~\eqref{eq:Pm1thermal} for a thermal state with temperature chosen to make $\bar n = \bar n_p$ in Eq.~\eqref{eq:barnr} for a chosen detuning of~$\Delta = \nu_p$.

An analogous calculation for the second term of~$P_{mn}^{(2)}$ shows that
\begin{multline}
\label{eq:Pmn2term2squeezed}
	\text{(Term~2)} =
	T^4 (\Omega_0 \eta)^4 \Biggl[ \sum_a \frac {b_m^{(p)} b_n^{(p)}} {\sqrt {\mu_p}} \Bigl\{ (\bar n + 1)  \sinc^2 \bigl[ (\Delta + \nu_p) \tfrac T 2 \bigr] + \bar n \sinc^2 \bigl[ (\Delta - \nu_p) \tfrac T 2 \bigr] \\
		 - 2\sqrt{\bar n (\bar n + 1)} \sinc \bigl[ (\Delta - \nu_p) \tfrac T 2 \bigr]  \sinc \bigl[ (\Delta + \nu_p) \tfrac T 2 \bigr]  \Bigr\} \Biggr]^2\;.
\end{multline}
In the limit of~$T \gg \nu^{-1}$, this gives no change from the same term in the equivalent thermal case:
\begin{align}
	\text{(Term~2)}_{\text{squeezed}} \xrightarrow{\text{($T \gg \nu^{-1}$)}} \text{(Term~2)}_{\text{thermal}}\;,
\end{align}
In the thermal case, the third term vanished in the long detection time-limit.  In the squeezed case, it does not, and this generates the difference in the correlation functions:
\begin{align}
\label{eq:eq:Pmnterm3squeezedlim}
	\text{(Term~3)} \xrightarrow{\text{($T \gg \nu^{-1}$)}} 
	T^4 (\Omega_0 \eta)^4 \left[ \sum_a \frac {b_m^{(a)} b_n^{(a)}} {\sqrt {\mu_a}} \sqrt{\bar n (\bar n + 1)} \Bigl\{ \delta_{(\Delta + \nu_a)} + \delta_{(\Delta - \nu_a)} \Bigr\} \right]^2\;.
\end{align}
This shows that even though local measurements have the same excitation statistics as the equivalent thermal state, simultaneous measurements of two ions do not---the correlations are stronger in the squeezed case, as should be expected.  For the uniformly squeezed state, we have
\begin{align}
\label{eq:fsqueezed}
	f_{mn} \simeq \frac {P_{mn}^{(2)}} {P_m^{(1)} P_n^{(1)}} = 3 + \frac {1} {\bar n}\;,
\end{align}
for any detuning~$\Delta = \pm \nu_p$ and any choice of ions~$(m,n)$ to measure.  The value depends on the squeezing parameter through~$\bar n$.  The reason the two do not agree in the limit~$\bar n \to 0$ is that in that case~$P_m = P_n = 0$, so $f_{mn}$~is undefined.  The behavior of this function, too, is consistent with the expected results for a second-order correlation function for such a squeezed state~\cite{Walls2008}.

\section{Discussion and Conclusion}
\label{sec:Discussion}

\begin{figure}
\label{fig:PmPn}
\includegraphics[width=.9 \columnwidth]{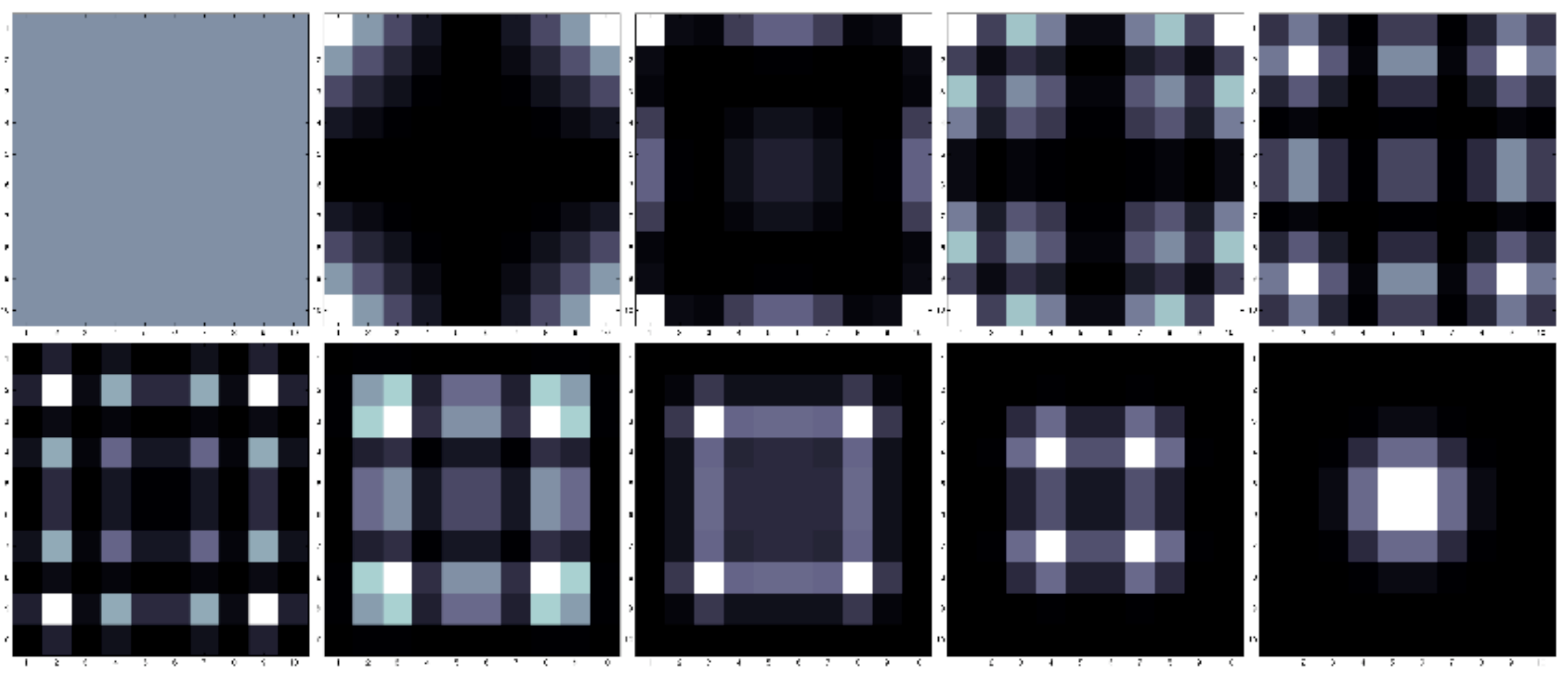}
\caption{Correlation functions~$P_{mn}$ for the normal modes of 10 ions in the case of the examples considered in Section~\ref{sec:examples}.  The top row represents~$P_{mn}$ for a detuning corresponding to normal modes~1 through~5; the bottom row, modes~6 through~10.  Since~$P_{mn} \propto P_m P_n$ for both types of states, except in the case of the center-of-mass mode (mode~1), which is uniform anyway, the colors are normalized such that white represents the maximum value, and black represents the minimum in each case.  Thermal and uniformly squeezed states both give the same results under this normalization.  The diagrams show two things: (1)~that the structure of the modes revealed by the correlations, and (2)~which pairs of ions (white boxes) are most useful for measurement when detuning to a given mode frequency.}
\label{fig:corr}
\end{figure}

There are really two pieces of information that can be gleaned from the correlation function~$P_{mn}$.  The first is the structure of the normal modes.  This structure can be probed by detuning the interaction laser to the desired mode's resonant frequency~$\nu_p$ and reading out ion~$m$ and~$n$.  Doing this for either the thermal state or the uniformly squeezed state from the previous section give correlations as in Figure~\ref{fig:corr}.
%------------------------
%------------------------

The other piece of information obtainable from~$P_{mn}$ is its relation to the single-ion detection probabilities~$P_m$ and~$P_n$.  This information is embodied in the normalized function~$f_{mn}$, and can be used to probe more general characteristics about the state.  For instance, in the comparison of squeezed and equivalent thermal states, with equivalence defined as an equal average number of phonons in the mode being detected (i.e.,~$\bar n = \bar n_p$ for~$\Delta = \nu_p$), many of the parameters cease to be important (detection time, coupling constant, geometric terms, etc.), and information about the basic nature of the state---in this case, thermal or squeezed---is revealed.

There is still much work to be done in exploring properties of correlations in a string of trapped ions.  The work presented here is designed to be a significant start in that direction.  We began with a definition of a measurable correlation function and single-ion detection probabilities and connected these to two- and four-point functions of a general quantum state, all in the perturbative regime.  Next, we specialized to Gaussian states and provided explicit formulas for these probabilities in terms of the state's covariance matrix elements.  The examples of thermal and squeezed states were compared, and a normalized correlation signature contrasted in each case, which shows agreement with standard results for correlation functions used in quantum optics.  Open problems include vast opportunities to generalize these results to other interesting states, such number states, coherent states, superpositions versus mixed states, and many others.  The structure of a string of trapped ions approximates a scalar field, but the deviations from this approximation could be a source of interest, as well.  Finally, a completely unexplored avenue would be to look at the behavior of these correlation functions as the trap frequency is modulated, to see whether such a modulation has a detectable correlation signature.

%%%%%%%%%%%%%%%%%%%%%%%%%%%%%%%%%%
% Acknowledgments
%%%%%%%%%%%%%%%%%%%%%%%%%%%%%%%%%%

\acknowledgments

We thank Steve Flammia and Andrew Doherty for useful discussions.  NCM was supported by the United States Department of Defense and the National Science Foundation.

\appendix*

%%%%%%%%%%%%%%%%%%%%%%%%%%%%%%%%%%
% Wick's Theorem
%%%%%%%%%%%%%%%%%%%%%%%%%%%%%%%%%%

\section{Wick's Theorem}
\label{sec:wick}

Wick's theorem theorem is often stated as a result for time-ordered expectation values of the ground state of a multimode bosonic or fermionic system, as in quantum field theory~\cite{Peskin1995}, but it can actually be used with any state and any prescribed ordering of the operators.  Wick's theorem is commonly stated as~\cite{Peskin1995}:
\begin{align}
\label{eq:Wickusual}
	\timeord\{\hat \phi_{m_1}(t_1)\dotsm \hat \phi_{m_N}(t_N)\} = \lcolon \hat \phi_{m_1}(t_1)\dotsm \hat \phi_{m_N}(t_N) \rcolon + \lcolon \text{(all possible contractions)} \rcolon\;,
\end{align}
where the colons in $\lcolon \text{(operators)} \rcolon$ place the contained operators in normal order (i.e., all raising operators are to the left of all lowering operators), a ``contraction'' of two operators is written as (and defined by)
\begin{align}
\label{eq:Tcontract}
	\acontraction[1ex]{}{\hat \phi}{_{m}(t_j)}{\hat \phi}
	\hat \phi_{m}(t_j) \hat \phi_{n} (t_k) \coloneqq 
	\begin{cases}
		[ \hat \psi_m(t_j), \hat \psi^\dag_n(t_k) ] & \text{if $t_j > t_k$}\;,\\
		[ \hat \psi_n(t_k), \hat \psi^\dag_m(t_j) ] & \text{if $t_j < t_k$}\;,
	\end{cases}
\end{align}
where~$\psi_m(t)$ and~$\psi^\dag_m(t)$ are defined in Eq.~\eqref{eq:ionposnegfreq}, and ``all possible contractions'' means every unique way of contracting pairs of operators together in this fashion.

We can make two generalizations: to anti-time-ordered products and to products with no time ordering.  Since normal ordering already prescribes and order for the uncontracted operators, the only place time ordering shows up at all is in the contracted terms.  As such, we define a second type of contraction for this purpose:
\begin{align}
\label{eq:Tbarcontract}
	\bcontraction[1ex]{}{\hat \phi}{_{m}(t_j)}{\hat \phi}
	\hat \phi_{m}(t_j) \hat \phi_{n} (t_k) \coloneqq 
	\begin{cases}
		[ \hat \psi_m(t_j), \hat \psi^\dag_n(t_k) ] & \text{if $t_j < t_k$}\;,\\
		[ \hat \psi_n(t_k), \hat \psi^\dag_m(t_j) ] & \text{if $t_j > t_k$}\;.
	\end{cases}
\end{align}
Finally, if the operator order is not prescribed by either type of time ordering (and is to be taken as given), then we shall define
\begin{align}
\label{eq:contract}
	\acontraction[1ex]{}{\hat \phi}{_{m}(t_j)}{\hat \phi}
	\bcontraction[1ex]{}{\hat \phi}{_{m}(t_j)}{\hat \phi}
	\hat \phi_{m}(t_j) \hat \phi_{n} (t_k) \coloneqq [ \hat \psi_m(t_j), \hat \psi^\dag_n(t_k) ]\;.
\end{align}
Generalizing the usual inductive method of Ref.~\cite{Peskin1995}, it's straighforward to generalize Wick's theorem to the following:
\begin{align}
\label{eq:Wickgeneral}
	\mathcal P \{\hat \phi_{m_1}(t_1)\dotsm \hat \phi_{m_N}(t_N)\} = \lcolon \hat \phi_{m_1}(t_1)\dotsm \hat \phi_{m_N}(t_N) \rcolon + \lcolon \text{(all possible contractions)} \rcolon\;,
\end{align}
where $\mathcal P\{ \dotsm \}$ prescribes a time ordering for (some of) the operators within it, and the contraction terms now must respect this ordering, using one of the above definitions in each case according to the prescribed ordering of the two operators involved.

For our purposes, we are interested in the four-point function in Eq.~\eqref{eq:Pmn2calc}.  This also serves as a particularly good but simple example of how to use Eq.~\eqref{eq:Wickgeneral}.  We'll use the following abbreviations to save space:
\begin{align}
\label{eq:qabbrev}
	\hat \phi_n(t_1) \to \hat \phi^1\;, \quad \hat \phi_m(t_2) \to \hat \phi^2\;, \quad \hat \phi_m(t_3) \to \hat \phi^3\;, \quad \hat \phi_n(t_4) \to \hat \phi^4\;.
\end{align}
We then can use Eq.~\eqref{eq:Wickgeneral} to make the following expansion:
\begin{align}
\label{eq:fourpointWickexpand0}
	\bar \timeord \{ \hat \phi^2 \hat \phi^1 \} \timeord \{ \hat \phi^3 \hat \phi^4 \} &= \lcolon \hat \phi^2 \hat \phi^1 \hat \phi^3 \hat \phi^4 \rcolon
	\nonumber \\ & \quad
	+ \lcolon
		\acontraction[1ex]{}{q}{^2 \hat \phi^1}{q}
		\bcontraction[1ex]{}{q}{^2 \hat \phi^1}{q}
		\hat \phi^2 \hat \phi^1 \hat \phi^3 \hat \phi^4 \rcolon
	+ \lcolon
		\acontraction[1ex]{\hat \phi^2}{q}{^1 \hat \phi^3}{q}
		\bcontraction[1ex]{\hat \phi^2}{q}{^1 \hat \phi^3}{q}
		\hat \phi^2 \hat \phi^1 \hat \phi^3 \hat \phi^4 \rcolon
	+ \lcolon
		\acontraction[1ex]{}{q}{^2 \hat \phi^1}{q}
		\bcontraction[1ex]{}{q}{^2 \hat \phi^1}{q}
		\acontraction[1.5ex]{\hat \phi^2}{q}{^1 \hat \phi^3}{q}
		\bcontraction[1.5ex]{\hat \phi^2}{q}{^1 \hat \phi^3}{q}
		\hat \phi^2 \hat \phi^1 \hat \phi^3 \hat \phi^4 \rcolon
	\nonumber \\ & \quad
	+ \lcolon
		\acontraction[1ex]{}{q}{^2 \hat \phi^1 \hat \phi^3}{q}
		\bcontraction[1ex]{}{q}{^2 \hat \phi^1 \hat \phi^3}{q}
		\hat \phi^2 \hat \phi^1 \hat \phi^3 \hat \phi^4 \rcolon
	+ \lcolon
		\acontraction[1ex]{\hat \phi^2}{q}{^1}{q}
		\bcontraction[1ex]{\hat \phi^2}{q}{^1}{q}
		\hat \phi^2 \hat \phi^1 \hat \phi^3 \hat \phi^4 \rcolon
	+ \lcolon
		\acontraction[1.5ex]{}{q}{^2 \hat \phi^1 \hat \phi^3}{q}
		\bcontraction[1.5ex]{}{q}{^2 \hat \phi^1 \hat \phi^3}{q}
		\acontraction[1ex]{\hat \phi^2}{q}{^1}{q}
		\bcontraction[1ex]{\hat \phi^2}{q}{^1}{q}
		\hat \phi^2 \hat \phi^1 \hat \phi^3 \hat \phi^4 \rcolon
	\nonumber \\ & \quad
	+ \lcolon
		\bcontraction[1ex]{}{q}{^2}{q}
		\hat \phi^2 \hat \phi^1 \hat \phi^3 \hat \phi^4 \rcolon
	+ \lcolon
		\acontraction[1ex]{\hat \phi^2 \hat \phi^1}{q}{^3}{q}
		\hat \phi^2 \hat \phi^1 \hat \phi^3 \hat \phi^4 \rcolon
	+ \lcolon
		\bcontraction[1ex]{}{q}{^2}{q}
		\acontraction[1ex]{\hat \phi^2 \hat \phi^1}{q}{^3}{q}
		\hat \phi^2 \hat \phi^1 \hat \phi^3 \hat \phi^4 \rcolon
	\;.
\end{align}
Notice how the contraction type is determined by whether the terms are prescribed to be in time order ($\hat \phi^3$ and~$\hat \phi^4$), anti-time order ($\hat \phi^2$ and $\hat \phi^1$), or as written (all other combinations).  We need the expectation value of each of these terms.  Since contractions of position operators are c-numbers, they can be taken out of the normal ordering and out of all expectation values, giving
\begin{align}
\label{eq:fourpointWickexpand}
	\langle \bar \timeord \{ \hat \phi^2 \hat \phi^1 \} \timeord \{ \hat \phi^3 \hat \phi^4 \} \rangle &= \langle \lcolon \hat \phi^2 \hat \phi^1 \hat \phi^3 \hat \phi^4 \rcolon \rangle
	\nonumber \\ & \quad
	+
		\acontraction[1ex]{}{q}{^2}{q}
		\bcontraction[1ex]{}{q}{^2}{q}
		\hat \phi^2 \hat \phi^3 \langle \lcolon \hat \phi^1 \hat \phi^4 \rcolon \rangle
	+
		\langle \lcolon \hat \phi^2 \hat \phi^3 \rcolon \rangle
		\acontraction[1ex]{}{q}{^1}{q}
		\bcontraction[1ex]{}{q}{^1}{q}
		\hat \phi^1 \hat \phi^4
	+
		\acontraction[1ex]{}{q}{^2}{q}
		\bcontraction[1ex]{}{q}{^2}{q}
		\acontraction[1ex]{\hat \phi^2 \hat \phi^3}{q}{^1}{q}
		\bcontraction[1ex]{\hat \phi^2 \hat \phi^3}{q}{^1}{q}
		\hat \phi^2 \hat \phi^3 \hat \phi^1 \hat \phi^4
	\nonumber \\ & \quad
	+
		\acontraction[1ex]{}{q}{^2}{q}
		\bcontraction[1ex]{}{q}{^2}{q}
		\hat \phi^2 \hat \phi^4 \langle \lcolon \hat \phi^1 \hat \phi^3 \rcolon \rangle
	+
		\langle \lcolon \hat \phi^2 \hat \phi^4 \rcolon \rangle
		\acontraction[1ex]{}{q}{^1}{q}
		\bcontraction[1ex]{}{q}{^1}{q}
		\hat \phi^1 \hat \phi^3
	+
		\acontraction[1ex]{}{q}{^2}{q}
		\bcontraction[1ex]{}{q}{^2}{q}
		\acontraction[1ex]{\hat \phi^2 \hat \phi^4}{q}{^1}{q}
		\bcontraction[1ex]{\hat \phi^2 \hat \phi^4}{q}{^1}{q}
		\hat \phi^2 \hat \phi^4 \hat \phi^1 \hat \phi^3
	\nonumber \\ & \quad
	+
		\bcontraction[1ex]{}{q}{^2}{q}
		\hat \phi^2 \hat \phi^1 \langle \lcolon \hat \phi^3 \hat \phi^4 \rcolon \rangle
	+
		\langle \lcolon \hat \phi^2 \hat \phi^1 \rcolon \rangle
		\acontraction[1ex]{}{q}{^3}{q}
		\hat \phi^3 \hat \phi^4
	+
		\bcontraction[1ex]{}{q}{^2}{q}
		\acontraction[1ex]{\hat \phi^2 \hat \phi^1}{q}{^3}{q}
		\hat \phi^2 \hat \phi^1 \hat \phi^3 \hat \phi^4
	\;.
\end{align}
The grouping of the terms in Eq.~\eqref{eq:fourpointWickexpand} suggests an alternate way of writing the expectation value of this expression.  Using Eq.~\eqref{eq:Wickgeneral}, we have
% NOTE: these labels suck
\begin{align}
\label{eq:double2pt2}
	\langle \hat \phi^2 \hat \phi^3 \rangle \langle \hat \phi^1 \hat \phi^4 \rangle
	&= \langle \lcolon \hat \phi^2 \hat \phi^3 \rcolon \rangle \langle \lcolon \hat \phi^1 \hat \phi^4 \rcolon \rangle
	+
		\acontraction[1ex]{}{q}{^2}{q}
		\bcontraction[1ex]{}{q}{^2}{q}
		\hat \phi^2 \hat \phi^3 \langle \lcolon \hat \phi^1 \hat \phi^4 \rcolon \rangle
	+
		\langle \lcolon \hat \phi^2 \hat \phi^3 \rcolon \rangle
		\acontraction[1ex]{}{q}{^1}{q}
		\bcontraction[1ex]{}{q}{^1}{q}
		\hat \phi^1 \hat \phi^4
	+
		\acontraction[1ex]{}{q}{^2}{q}
		\bcontraction[1ex]{}{q}{^2}{q}
		\acontraction[1ex]{\hat \phi^2 \hat \phi^3}{q}{^1}{q}
		\bcontraction[1ex]{\hat \phi^2 \hat \phi^3}{q}{^1}{q}
		\hat \phi^2 \hat \phi^3 \hat \phi^1 \hat \phi^4
	\;,\\
\label{eq:double2pt3}
	\langle \hat \phi^2 \hat \phi^4 \rangle \langle \hat \phi^1 \hat \phi^3 \rangle
	&= \langle \lcolon \hat \phi^2 \hat \phi^4 \rcolon \rangle \langle \lcolon \hat \phi^1 \hat \phi^3 \rcolon \rangle
	+
		\acontraction[1ex]{}{q}{^2}{q}
		\bcontraction[1ex]{}{q}{^2}{q}
		\hat \phi^2 \hat \phi^4 \langle \lcolon \hat \phi^1 \hat \phi^3 \rcolon \rangle
	+
		\langle \lcolon \hat \phi^2 \hat \phi^4 \rcolon \rangle
		\acontraction[1ex]{}{q}{^1}{q}
		\bcontraction[1ex]{}{q}{^1}{q}
		\hat \phi^1 \hat \phi^3
	+
		\acontraction[1ex]{}{q}{^2}{q}
		\bcontraction[1ex]{}{q}{^2}{q}
		\acontraction[1ex]{\hat \phi^2 \hat \phi^4}{q}{^1}{q}
		\bcontraction[1ex]{\hat \phi^2 \hat \phi^4}{q}{^1}{q}
		\hat \phi^2 \hat \phi^4 \hat \phi^1 \hat \phi^3
	\;, \\
\label{eq:double2pt1}
	\langle \bar \timeord \{ \hat \phi^2 \hat \phi^1 \} \rangle \langle \timeord \{ \hat \phi^3 \hat \phi^4 \} \rangle
	&= \langle \lcolon \hat \phi^2 \hat \phi^1 \rcolon \rangle \langle \lcolon \hat \phi^3 \hat \phi^4 \rcolon \rangle
	+
		\bcontraction[1ex]{}{q}{^2}{q}
		\hat \phi^2 \hat \phi^1 \langle \lcolon \hat \phi^3 \hat \phi^4 \rcolon \rangle
	+
		\langle \lcolon \hat \phi^2 \hat \phi^1 \rcolon \rangle
		\acontraction[1ex]{}{q}{^3}{q}
		\hat \phi^3 \hat \phi^4
	+
		\bcontraction[1ex]{}{q}{^2}{q}
		\acontraction[1ex]{\hat \phi^2 \hat \phi^1}{q}{^3}{q}
		\hat \phi^2 \hat \phi^1 \hat \phi^3 \hat \phi^4
	\;,
\end{align}
Using this, we can write
\begin{multline}
\label{eq:fourpointWickexpandalt}
	\langle \bar \timeord \{ \hat \phi^2 \hat \phi^1 \} \timeord \{ \hat \phi^3 \hat \phi^4 \} \rangle = \langle \hat \phi^2 \hat \phi^3 \rangle \langle \hat \phi^1 \hat \phi^4 \rangle + \langle \hat \phi^2 \hat \phi^4 \rangle \langle \hat \phi^1 \hat \phi^3 \rangle + \langle \bar \timeord \{ \hat \phi^2 \hat \phi^1 \} \rangle \langle \timeord \{ \hat \phi^3 \hat \phi^4 \} \rangle
	\\
	+ \langle \lcolon \hat \phi^2 \hat \phi^1 \hat \phi^3 \hat \phi^4 \rcolon \rangle - \langle \lcolon \hat \phi^2 \hat \phi^3 \rcolon \rangle \langle \lcolon \hat \phi^1 \hat \phi^4 \rcolon \rangle - \langle \lcolon \hat \phi^2 \hat \phi^4 \rcolon \rangle \langle \lcolon \hat \phi^1 \hat \phi^3 \rcolon \rangle - \langle \lcolon \hat \phi^2 \hat \phi^1 \rcolon \rangle \langle \lcolon \hat \phi^3 \hat \phi^4 \rcolon \rangle \;.
\end{multline}
Recalling the abbreviations~\eqref{eq:qabbrev}, either Eq.~\eqref{eq:fourpointWickexpand} or Eq.~\eqref{eq:fourpointWickexpandalt} may be used to calculate $P_{mn}$ to lowest nontrivial order.

% 
% Gaussian states
% 

\subsection*{Gaussian States}
%\label{subsec:gaussian}

If the state has a Wigner function that is a Gaussian with zero mean, then all of the normal-ordered terms in Eq.~\eqref{eq:fourpointWickexpandalt} must cancel out in order to agree with the result known as the ``generalized Wick's theorem'' from Ref.~\cite{Louisell1990}.  As it is stated in Ref.~\cite{Louisell1990}, the theorem applies only to ground-state expectation values of raising and lowering operators, but it can be generalized to a much larger class of expectation values.  Let $\hat \kappa_i \in \{ \hat a_p, \hat a_p^\dag \mid p = 1,\dotsc,N \}$ be any raising or lowering operator.  The generalized Wick's theorem states that any ground-state expectation value of an even number of such operators can be written in the following form:
\begin{align}
\label{eq:Wickvacuum}
	\bra 0 \hat \kappa_1 \hat \kappa_2 \hat \kappa_3 \dotsm \hat \kappa_{2n} \ket 0 = \sum_{P_d} \bra 0 \hat \kappa_1 \hat \kappa_2 \ket 0 \bra 0 \hat \kappa_3 \hat \kappa_4 \ket 0 \dotsm \bra 0 \hat \kappa_{2n-1} \hat \kappa_{2n} \ket 0\;, 
\end{align}
where the sum is over all {\it distinct} permutations $P_d$ of the $2n$ indices that preserve the operator ordering within each pairing on the right---i.e., all permutations which give a distinct product of expectation-value pairs $\langle \hat \kappa_l \hat \kappa_m \rangle$ and for which $l < m$ in each of these pairs.  Also, any linear functions $\hat F_i = \sum_i A_{ij} \hat \kappa_j$ of the raising and lowering operators will separate in a similar fashion:
\begin{align}
\label{eq:Wicklinearvacuum}
	\bra 0 \hat F_1 \hat F_2 \hat F_3 \dotsm \hat F_{2n} \ket 0 = \sum_{P_d} \bra 0 \hat F_1 \hat F_2 \ket 0 \bra 0 \hat F_3 \hat F_4 \ket 0 \dotsm \bra 0 \hat F_{2n-1} \hat F_{2n} \ket 0\;,
\end{align}
as can be verified by direct calculation.  Furthermore, it turns out that this property holds for all zero-mean Gaussian states, both pure and mixed.  To show this, one notes that any such Gaussian state can be written as the partial trace of a pure Gaussian state of twice as many modes~\cite{Holevo2001}: $\rho = \tr_B \outprod \chi \chi$.  Such a Gaussian pure state $\ket \chi$ is unitarily related to the vacuum state on the doubled mode set---i.e., $\ket \chi = U_\chi \ket {0,0}$.  However, this unitary transformation can be viewed in the Heisenberg picture as a symplectic linear transformation on the (extended) vector of canonical operators~\cite{Bartlett2002}, viz.\ $U_\chi^\dag \hat \kappa_j U_\chi = \sum_k L^\chi_{jk} \hat \varphi_k =: \hat G_j$, where $\hat \kappa_j$ is any raising or lowering operator in the original set, while $\hat \varphi_k$ is any such operator in the extended set.  From this, we can write expectation values as
\begin{align}
\label{eq:psijprod}
	\langle \hat \kappa_1 \dotsm \hat \kappa_n \rangle &= \tr ( \rho \hat \kappa_1 \dotsm \hat \kappa_n )  \nonumber \\
	&= \tr_A \bigl[ \tr_B (\outprod \chi \chi ) \hat \kappa_1 \hat \kappa_2 \hat \kappa_3 \dotsm \hat \kappa_{2n} \bigr] \nonumber \\
	&= \bra \chi \hat \kappa_1 \dotsm \hat \kappa_n \ket \chi \nonumber \\
	&= \bra {0,0} U_\chi^\dag \hat \kappa_1 U_\chi \dotsm U_\chi^\dag \hat \kappa_n U_\chi \ket {0,0} \nonumber \\
	&= \bra {0,0} \hat G_1 \dotsm \hat G_n \ket {0,0}\;.
\end{align}
Thus, we have
\begin{align}
\label{eq:Wick}
	\langle \hat \kappa_1 \hat \kappa_2 \hat \kappa_3 \dotsm \hat \kappa_{2n} \rangle 
	&= \bra {0,0} \hat G_1 \hat G_2 \hat G_3 \dotsm \hat G_{2n} \ket {0,0} \nonumber  \\
	&= \bra {0,0} \hat G_1 \hat G_2 \ket {0,0} \bra {0,0} \hat G_3 \hat G_4 \ket {0,0} \dotsm \bra {0,0} \hat G_{2n-1} \hat G_{2n} \ket {0,0} \nonumber \\
	&= \langle \hat \kappa_1 \hat \kappa_2 \rangle \langle \hat \kappa_3 \hat \kappa_4 \rangle \dotsm \langle \hat \kappa_{2n-1} \hat \kappa_{2n} \rangle\;,
\end{align}
where we applied Eq.~\eqref{eq:Wicklinearvacuum} in the middle step.  Thus, the generalized Wick's theorem holds for all Gaussian states with zero mean.  Furthermore, we may take linear combinations of $\hat \kappa_j$ operators, and the theorem still holds:
\begin{align}
\label{eq:Wicklinear}
	\langle \hat F_1 \hat F_2 \hat F_3 \dotsm \hat F_{2n} \rangle = \sum_{P_d} \langle \hat F_1 \hat F_2 \rangle \langle \hat F_3 \hat F_4 \rangle \dotsm \langle \hat F_{2n-1} \hat F_{2n} \rangle\;.
\end{align}
If the state in Eq.~\eqref{eq:fourpointWickexpandalt} is a zero-mean Gaussian, then Eq.~\eqref{eq:Wicklinear} applies, resulting in only the first three terms on the right being kept:
\begin{multline}
\label{eq:Pmn3terms}
	\langle \bar \timeord \{ \hat \phi_m(t_2) \hat \phi_n(t_1) \} \timeord \{ \hat \phi_m(t_3) \hat \phi_n(t_4) \} \rangle \xrightarrow[\text{state}]{\text{Gaussian}} \langle \hat \phi_m(t_2) \hat \phi_m(t_3) \rangle \langle \hat \phi_n(t_1) \hat \phi_n(t_4) \rangle \\
	+ \langle \hat \phi_m(t_2) \hat \phi_n(t_4) \rangle \langle \hat \phi_n(t_1) \hat \phi_m(t_3) \rangle 
	+ \langle \bar \timeord \{ \hat \phi_m(t_2) \hat \phi_n(t_1) \} \rangle \langle \timeord \{ \hat \phi_m(t_3) \hat \phi_n(t_4) \} \rangle\;.
\end{multline}
Comparing this with Eq.~\eqref{eq:Pm1calc}, we see that the first term will always give simply $P_m P_n$, and the second will always give a term similar to this, but with a different geometric factor.  Thus, we need only ever explicitly calculate the last two terms from Eq.~\eqref{eq:Pmn3terms}.

We also note that if we restrict to Gaussian vibrational states, the generalized Wick's theorem says that all higher-order joint probabilities (those with more excited ions, like $P_{lmn}$, and/or higher-order calculations) will all decompose into integrals and sums of expectation values of the form $\langle \hat \phi_m(t_1) \hat \phi_n(t_2) \rangle$ and $\langle \timeord \{ \hat \phi_m(t_1) \hat \phi_n(t_2) \} \rangle$ for arbitrary $m$ and $n$.  (Note that the antitime-ordered case is just the complex conjugate of the time-ordered case).  When restricting to Gaussian states, then, only these two types of expectation values are required.

%%%%%%%%%%%%%%%%%%%%%%%%%%%%%%%%%%
% Bibliography
%%%%%%%%%%%%%%%%%%%%%%%%%%%%%%%%%%

\bibliography{IonCorrelations}

\begin{thebibliography}{25}
\expandafter\ifx\csname natexlab\endcsname\relax\def\natexlab#1{#1}\fi
\expandafter\ifx\csname bibnamefont\endcsname\relax
  \def\bibnamefont#1{#1}\fi
\expandafter\ifx\csname bibfnamefont\endcsname\relax
  \def\bibfnamefont#1{#1}\fi
\expandafter\ifx\csname citenamefont\endcsname\relax
  \def\citenamefont#1{#1}\fi
\expandafter\ifx\csname url\endcsname\relax
  \def\url#1{\texttt{#1}}\fi
\expandafter\ifx\csname urlprefix\endcsname\relax\def\urlprefix{URL }\fi
\providecommand{\bibinfo}[2]{#2}
\providecommand{\eprint}[2][]{\url{#2}}

\bibitem[{\citenamefont{Bernui}(2005)}]{Bernui2005}
\bibinfo{author}{\bibfnamefont{A.}~\bibnamefont{Bernui}},
  \bibinfo{journal}{Braz.\ J.\ Phys.} \textbf{\bibinfo{volume}{35}},
  \bibinfo{pages}{1185} (\bibinfo{year}{2005}).

\bibitem[{\citenamefont{Grishchuk}(1996)}]{Grishchuk1996}
\bibinfo{author}{\bibfnamefont{L.~P.} \bibnamefont{Grishchuk}},
  \bibinfo{journal}{Phys. Rev. D} \textbf{\bibinfo{volume}{53}},
  \bibinfo{pages}{6784} (\bibinfo{year}{1996}).

\bibitem[{\citenamefont{Wen}(2004)}]{Wen2004}
\bibinfo{author}{\bibfnamefont{X.-G.} \bibnamefont{Wen}},
  \emph{\bibinfo{title}{Quantum Field Theorey of Many-Body Systems}}
  (\bibinfo{address}{Oxford},
  \bibinfo{year}{2004}).

\bibitem[{\citenamefont{Posazhennikova}(2006)}]{Posazhennikova2006}
\bibinfo{author}{\bibfnamefont{A.}~\bibnamefont{Posazhennikova}},
  \bibinfo{journal}{Rev.\ Mod.\ Phys.} \textbf{\bibinfo{volume}{78}},
  \bibinfo{eid}{1111} (\bibinfo{year}{2006}).

\bibitem[{\citenamefont{Hellweg et~al.}(2003)\citenamefont{Hellweg,
  Cacciapuoti, Kottke, Schulte, Sengstock, Ertmer, and Arlt}}]{Hellweg2003}
\bibinfo{author}{\bibfnamefont{D.}~\bibnamefont{Hellweg}},
  \bibinfo{author}{\bibfnamefont{L.}~\bibnamefont{Cacciapuoti}},
  \bibinfo{author}{\bibfnamefont{M.}~\bibnamefont{Kottke}},
  \bibinfo{author}{\bibfnamefont{T.}~\bibnamefont{Schulte}},
  \bibinfo{author}{\bibfnamefont{K.}~\bibnamefont{Sengstock}},
  \bibinfo{author}{\bibfnamefont{W.}~\bibnamefont{Ertmer}}, \bibnamefont{and}
  \bibinfo{author}{\bibfnamefont{J.~J.} \bibnamefont{Arlt}},
  \bibinfo{journal}{Phys. Rev. Lett.} \textbf{\bibinfo{volume}{91}},
  \bibinfo{pages}{010406} (\bibinfo{year}{2003}).

\bibitem[{\citenamefont{Burt et~al.}(1997)\citenamefont{Burt, Ghrist, Myatt,
  Holland, Cornell, and Wieman}}]{Burt1997}
\bibinfo{author}{\bibfnamefont{E.~A.} \bibnamefont{Burt}},
  \bibinfo{author}{\bibfnamefont{R.~W.} \bibnamefont{Ghrist}},
  \bibinfo{author}{\bibfnamefont{C.~J.} \bibnamefont{Myatt}},
  \bibinfo{author}{\bibfnamefont{M.~J.} \bibnamefont{Holland}},
  \bibinfo{author}{\bibfnamefont{E.~A.} \bibnamefont{Cornell}},
  \bibnamefont{and} \bibinfo{author}{\bibfnamefont{C.~E.}
  \bibnamefont{Wieman}}, \bibinfo{journal}{Phys. Rev. Lett.}
  \textbf{\bibinfo{volume}{79}}, \bibinfo{pages}{337} (\bibinfo{year}{1997}).

\bibitem[{\citenamefont{Kagan et~al.}(1985)\citenamefont{Kagan, Svistunov, and
  Shlyapnikov}}]{Kagan1985}
\bibinfo{author}{\bibfnamefont{Y.}~\bibnamefont{Kagan}},
  \bibinfo{author}{\bibfnamefont{B.~V.} \bibnamefont{Svistunov}},
  \bibnamefont{and} \bibinfo{author}{\bibfnamefont{G.~V.}
  \bibnamefont{Shlyapnikov}}, \bibinfo{journal}{Sov.\ J.\ Exp.\
  Theor.\ Phys.\ Lett.} \textbf{\bibinfo{volume}{42}},
  \bibinfo{pages}{209} (\bibinfo{year}{1985}).

\bibitem[{\citenamefont{Leibfried
  et~al.}(2003{\natexlab{a}})\citenamefont{Leibfried, Blatt, Monroe, and
  Wineland}}]{Leibfried2003}
\bibinfo{author}{\bibfnamefont{D.}~\bibnamefont{Leibfried}},
  \bibinfo{author}{\bibfnamefont{R.}~\bibnamefont{Blatt}},
  \bibinfo{author}{\bibfnamefont{C.}~\bibnamefont{Monroe}}, \bibnamefont{and}
  \bibinfo{author}{\bibfnamefont{D.}~\bibnamefont{Wineland}},
  \bibinfo{journal}{Rev. Mod. Phys.} \textbf{\bibinfo{volume}{75}},
  \bibinfo{pages}{281} (\bibinfo{year}{2003}{\natexlab{a}}).

\bibitem[{\citenamefont{Franke-Arnold et~al.}(2003)\citenamefont{Franke-Arnold,
  Barnett, Huyet, and Sailliot}}]{Franke-Arnold2003}
\bibinfo{author}{\bibfnamefont{S.}~\bibnamefont{Franke-Arnold}},
  \bibinfo{author}{\bibfnamefont{S.~M.} \bibnamefont{Barnett}},
  \bibinfo{author}{\bibfnamefont{G.}~\bibnamefont{Huyet}}, \bibnamefont{and}
  \bibinfo{author}{\bibfnamefont{C.}~\bibnamefont{Sailliot}},
  \bibinfo{journal}{Euro.\ Phys.\ D} \textbf{\bibinfo{volume}{22}},
  \bibinfo{pages}{373} (\bibinfo{year}{2003}).

\bibitem[{\citenamefont{Monroe et~al.}(1995)\citenamefont{Monroe, Meekhof,
  King, Jefferts, Itano, Wineland, and Gould}}]{Monroe1995}
\bibinfo{author}{\bibfnamefont{C.}~\bibnamefont{Monroe}},
  \bibinfo{author}{\bibfnamefont{D.~M.} \bibnamefont{Meekhof}},
  \bibinfo{author}{\bibfnamefont{B.~E.} \bibnamefont{King}},
  \bibinfo{author}{\bibfnamefont{S.~R.} \bibnamefont{Jefferts}},
  \bibinfo{author}{\bibfnamefont{W.~M.} \bibnamefont{Itano}},
  \bibinfo{author}{\bibfnamefont{D.~J.} \bibnamefont{Wineland}},
  \bibnamefont{and} \bibinfo{author}{\bibfnamefont{P.~L.} \bibnamefont{Gould}},
  \bibinfo{journal}{Phys. Rev. Lett.} \textbf{\bibinfo{volume}{75}},
  \bibinfo{pages}{4011} (\bibinfo{year}{1995}).

\bibitem[{\citenamefont{James}(1998)}]{James1998}
\bibinfo{author}{\bibfnamefont{D.~F.~V.} \bibnamefont{James}},
  \bibinfo{journal}{Appl.\ Phys.\ B}
  \textbf{\bibinfo{volume}{66}}, \bibinfo{pages}{181} (\bibinfo{year}{1998}).

\bibitem[{\citenamefont{Leibfried
  et~al.}(2003{\natexlab{b}})\citenamefont{Leibfried, De~Marco, Meyer, Lucas,
  Barrett, Britton, Itano, Jelenkovic, Langer, Rosenband
  et~al.}}]{Leibfried2003a}
\bibinfo{author}{\bibfnamefont{D.}~\bibnamefont{Leibfried}},
  \bibinfo{author}{\bibfnamefont{B.}~\bibnamefont{De~Marco}},
  \bibinfo{author}{\bibfnamefont{V.}~\bibnamefont{Meyer}},
  \bibinfo{author}{\bibfnamefont{D.}~\bibnamefont{Lucas}},
  \bibinfo{author}{\bibfnamefont{M.}~\bibnamefont{Barrett}},
  \bibinfo{author}{\bibfnamefont{J.}~\bibnamefont{Britton}},
  \bibinfo{author}{\bibfnamefont{W.~M.} \bibnamefont{Itano}},
  \bibinfo{author}{\bibfnamefont{B.}~\bibnamefont{Jelenkovic}},
  \bibinfo{author}{\bibfnamefont{C.}~\bibnamefont{Langer}},
  \bibinfo{author}{\bibfnamefont{T.}~\bibnamefont{Rosenband}},
  \bibnamefont{et~al.}, \bibinfo{journal}{Nature}
  \textbf{\bibinfo{volume}{422}}, \bibinfo{pages}{412}
  (\bibinfo{year}{2003}{\natexlab{b}}).

\bibitem[{\citenamefont{Schmidt-Kaler et~al.}(2003)\citenamefont{Schmidt-Kaler,
  H\"affner, Riebe, Lancaster, Deuschle, Becher, Roos, Eschner, and
  Blatt}}]{Schmidt-Kaler2003}
\bibinfo{author}{\bibfnamefont{F.}~\bibnamefont{Schmidt-Kaler}},
  \bibinfo{author}{\bibfnamefont{H.}~\bibnamefont{H\"affner}},
  \bibinfo{author}{\bibfnamefont{M.}~\bibnamefont{Riebe}},
  \bibinfo{author}{\bibfnamefont{G.~P.~T.} \bibnamefont{Lancaster}},
  \bibinfo{author}{\bibfnamefont{T.}~\bibnamefont{Deuschle}},
  \bibinfo{author}{\bibfnamefont{C.}~\bibnamefont{Becher}},
  \bibinfo{author}{\bibfnamefont{C.~F.} \bibnamefont{Roos}},
  \bibinfo{author}{\bibfnamefont{J.}~\bibnamefont{Eschner}}, \bibnamefont{and}
  \bibinfo{author}{\bibfnamefont{R.}~\bibnamefont{Blatt}},
  \bibinfo{journal}{Nature} \textbf{\bibinfo{volume}{422}},
  \bibinfo{pages}{408} (\bibinfo{year}{2003}).

\bibitem[{\citenamefont{Naraschewski and Glauber}(1999)}]{Naraschewski1999}
\bibinfo{author}{\bibfnamefont{M.}~\bibnamefont{Naraschewski}}
  \bibnamefont{and} \bibinfo{author}{\bibfnamefont{R.~J.}
  \bibnamefont{Glauber}}, \bibinfo{journal}{Phys. Rev. A}
  \textbf{\bibinfo{volume}{59}}, \bibinfo{pages}{4595} (\bibinfo{year}{1999}).

\bibitem[{\citenamefont{Kolobov}(1999)}]{Kolobov1999}
\bibinfo{author}{\bibfnamefont{M.~I.} \bibnamefont{Kolobov}},
  \bibinfo{journal}{Rev. Mod. Phys.} \textbf{\bibinfo{volume}{71}},
  \bibinfo{pages}{1539} (\bibinfo{year}{1999}).

\bibitem[{\citenamefont{Nogueira et~al.}(2001)\citenamefont{Nogueira, Walborn,
  P\'adua, and Monken}}]{Nogueira2001}
\bibinfo{author}{\bibfnamefont{W.~A.~T.} \bibnamefont{Nogueira}},
  \bibinfo{author}{\bibfnamefont{S.~P.} \bibnamefont{Walborn}},
  \bibinfo{author}{\bibfnamefont{S.}~\bibnamefont{P\'adua}}, \bibnamefont{and}
  \bibinfo{author}{\bibfnamefont{C.~H.} \bibnamefont{Monken}},
  \bibinfo{journal}{Phys. Rev. Lett.} \textbf{\bibinfo{volume}{86}},
  \bibinfo{pages}{4009} (\bibinfo{year}{2001}).

\bibitem[{\citenamefont{Kolobov and Fabre}(2000)}]{Kolobov2000}
\bibinfo{author}{\bibfnamefont{M.~I.} \bibnamefont{Kolobov}} \bibnamefont{and}
  \bibinfo{author}{\bibfnamefont{C.}~\bibnamefont{Fabre}},
  \bibinfo{journal}{Phys. Rev. Lett.} \textbf{\bibinfo{volume}{85}},
  \bibinfo{pages}{3789} (\bibinfo{year}{2000}).

\bibitem[{\citenamefont{Wallentowitz and Vogel}(1996)}]{Wallentowitz1996}
\bibinfo{author}{\bibfnamefont{S.}~\bibnamefont{Wallentowitz}}
  \bibnamefont{and} \bibinfo{author}{\bibfnamefont{W.}~\bibnamefont{Vogel}},
  \bibinfo{journal}{Phys. Rev. A} \textbf{\bibinfo{volume}{54}},
  \bibinfo{pages}{3322} (\bibinfo{year}{1996}).

\bibitem[{\citenamefont{Walls and Milburn}(2008)}]{Walls2008}
\bibinfo{author}{\bibfnamefont{D.~F.} \bibnamefont{Walls}} \bibnamefont{and}
  \bibinfo{author}{\bibfnamefont{G.~J.} \bibnamefont{Milburn}},
  \emph{\bibinfo{title}{Quantum Optics}} (\bibinfo{publisher}{Springer},
  \bibinfo{address}{Berlin}, \bibinfo{year}{2008}), \bibinfo{edition}{2nd} ed.

\bibitem[{\citenamefont{Jaynes and Cummings}(1963)}]{Jaynes1963}
\bibinfo{author}{\bibfnamefont{E.~T.} \bibnamefont{Jaynes}} \bibnamefont{and}
  \bibinfo{author}{\bibfnamefont{F.~W.} \bibnamefont{Cummings}},
  \bibinfo{journal}{Proceedings of the IEEE} \textbf{\bibinfo{volume}{51}},
  \bibinfo{pages}{89} (\bibinfo{year}{1963}).
  
\bibitem[{\citenamefont{DeWitt}(1979)}]{DeWitt1979}
\bibinfo{author}{\bibfnamefont{B.~S.} \bibnamefont{DeWitt}}, in
  \emph{\bibinfo{booktitle}{General Relativity, An Einstein Centenary Survey}}
  (\bibinfo{publisher}{Cambridge}, \bibinfo{year}{1979}).

\bibitem[{\citenamefont{Peskin and Schroeder}(1995)}]{Peskin1995}
\bibinfo{author}{\bibfnamefont{M.~E.} \bibnamefont{Peskin}} \bibnamefont{and}
  \bibinfo{author}{\bibfnamefont{D.~V.} \bibnamefont{Schroeder}},
  \emph{\bibinfo{title}{An Introduction to Quantum Field Theory}}
  (\bibinfo{publisher}{Perseus}, \bibinfo{address}{Cambridge, Massachusetts},
  \bibinfo{year}{1995}).

\bibitem[{\citenamefont{Louisell}(1990)}]{Louisell1990}
\bibinfo{author}{\bibfnamefont{W.~H.} \bibnamefont{Louisell}},
  \emph{\bibinfo{title}{Quantum Statistical Properties of Radiation}}
  (\bibinfo{publisher}{Wiley}, \bibinfo{address}{New York},
  \bibinfo{year}{1990}).

\bibitem[{\citenamefont{Holevo and Werner}(2001)}]{Holevo2001}
\bibinfo{author}{\bibfnamefont{A.~S.} \bibnamefont{Holevo}} \bibnamefont{and}
  \bibinfo{author}{\bibfnamefont{R.~F.} \bibnamefont{Werner}},
  \bibinfo{journal}{Phys. Rev. A} \textbf{\bibinfo{volume}{63}},
  \bibinfo{pages}{032312} (\bibinfo{year}{2001}).

\bibitem[{\citenamefont{Bartlett et~al.}(2002)\citenamefont{Bartlett, Sanders,
  Braunstein, and Nemoto}}]{Bartlett2002}
\bibinfo{author}{\bibfnamefont{S.~D.} \bibnamefont{Bartlett}},
  \bibinfo{author}{\bibfnamefont{B.~C.} \bibnamefont{Sanders}},
  \bibinfo{author}{\bibfnamefont{S.~L.} \bibnamefont{Braunstein}},
  \bibnamefont{and} \bibinfo{author}{\bibfnamefont{K.}~\bibnamefont{Nemoto}},
  \bibinfo{journal}{Phys. Rev. Lett.} \textbf{\bibinfo{volume}{88}},
  \bibinfo{pages}{097904}~(\bibinfo{year}{2002}).

\end{thebibliography}

\end{document}